\documentstyle[11pt,amssymb]{article}
\makeatletter
\@addtoreset{equation}{section}
\makeatother

\let\a=\alpha    
    
  \let\n=\nu

\let\C=\Chi

\def\nn{\nonumber} \def\bd{\begin{document}} \def\ed{\end{document}}
\def\ds{\documentstyle} \let\fr=\frac \let\bl=\bigl \let\br=\bigr
\let\Br=\Bigr \let\Bl=\Bigl 
\let\bm=\bibitem
\let\na=\nabla
\let\pa=\partial \let\ov=\overline 
\newcommand{\be}{\begin{equation}} 
\newcommand{\ee}{\end{equation}} 
\def\ba{\begin{array}}
\def\ea{\end{array}}
\def\ft#1#2{{\textstyle{{\scriptstyle #1}\over {\scriptstyle #2}}}}
\def\fft#1#2{{#1 \over #2}}
\def\del{\partial}
\def\vp{\varphi}
\def\st#1{{\scriptstyle #1}}
\def\sst#1{{\scriptscriptstyle #1}}

\def\oneone{\rlap 1\mkern4mu{\rm l}}
\def\td{\tilde}
\def\wtd{\widetilde}
\def\ie{\rm i.e.\ }
\def\dalemb#1#2{{\vbox{\hrule height .#2pt
        \hbox{\vrule width.#2pt height#1pt \kern#1pt
                \vrule width.#2pt}
        \hrule height.#2pt}}}
\def\square{\mathord{\dalemb{6.8}{7}\hbox{\hskip1pt}}}

\def\cramp{\medmuskip = 2mu plus 1mu minus 2mu}
\def\cramper{\medmuskip = 2mu plus 1mu minus 2mu}
\def\crampest{\medmuskip = 1mu plus 1mu minus 1mu}
\def\uncramp{\medmuskip = 4mu plus 2mu minus 4mu}

\newcommand{\ho}[1]{$\, ^{#1}$}
\newcommand{\hoch}[1]{$\, ^{#1}$}
\newcommand{\bea}{\begin{eqnarray}} 
\newcommand{\eea}{\end{eqnarray}} 
\newcommand{\ra}{\rightarrow}
\newcommand{\lra}{\longrightarrow}
\newcommand{\Lra}{\Leftrightarrow}
\newcommand{\ap}{\alpha^\prime}
\newcommand{\bp}{\tilde \beta^\prime}
\newcommand{\tr}{{\rm tr} }
\newcommand{\Tr}{{\rm Tr} } 
\def\0{{\sst{(0)}}}
\def\1{{\sst{(1)}}}
\def\2{{\sst{(2)}}}
\def\3{{\sst{(3)}}}
\def\4{{\sst{(4)}}}
\def\5{{\sst{(5)}}}
\def\6{{\sst{(6)}}}
\def\7{{\sst{(7)}}}
\def\8{{\sst{(8)}}}
\def\n{{\sst{(n)}}}
\def\cA{{{\cal A}}}
\def\cF{{{\cal F}}}
\def\tV{\widetilde V}
\def\tW{\widetilde W}
\def\tH{\widetilde H}
\def\tE{\widetilde E}
\def\tF{\widetilde F}
\def\tA{\widetilde A}
\def\im{{{\rm i}}}
\def\jm{{{\rm j}}}
\def\km{{{\rm k}}}

\def\tY{{{\wtd Y}}}
\def\ep{{\epsilon}}
\def\vep{{\varepsilon}}
\def\R{\rlap{\rm I}\mkern3mu{\rm R}}
\def\bD{{{\bar D}}}
\def\R{{{\Bbb R}}}
\def\C{{{\Bbb C}}}
\def\H{{{\Bbb H}}}
\def\CP{{{\Bbb C}{\Bbb P}}}
\def\RP{{{\Bbb R}{\Bbb P}}}
\def\Z{{{\Bbb Z}}}

\newcommand{\NP}{Nucl. Phys. }
\newcommand{\tamphys}{\it Center for Theoretical Physics\\
Texas A\&M University, College Station, TX 77843}
\newcommand{\umich}{\it Michigan Center for Theoretical Physics\\
University of Michigan, Ann Arbor, Michigan 48109}
\newcommand{\upenn}{\it Department of Physics and Astronomy\\
University of Pennsylvania, Philadelphia,  PA 19104}
\newcommand{\SISSA}{\it  SISSA-ISAS and INFN, Sezione di Trieste\\
Via Beirut 2-4, I-34013, Trieste, Italy}

\newcommand{\ihp}{\it Institut Henri Poincar\'e\\
  11 rue Pierre et Marie Curie, F 75231 Paris Cedex 05}

\newcommand{\damtp}{\it DAMTP, Centre for Mathematical Sciences,
 Cambridge University, Wilberforce Road, Cambridge CB3 OWA, UK}

\newcommand{\auth}{M. Cveti\v{c}\hoch{\dagger}, G.W. Gibbons\hoch{\sharp}, 
H. L\"u\hoch{\star} and C.N. Pope\hoch{\ddagger}}

\begin{document}
\begin{flushright}
\hfill{DAMTP-2001-19}\ \ \ {CTP TAMU-07/01}\ \ \ {UPR-928-T}\ \ \
{IHP-2000/22}\ \ \   {MCTP-01-10}\\ 
{February 2001}\ \ \
{hep-th/0102185}
\end{flushright}

%\vspace{15pt}

\begin{center}
{ \large {\bf Hyper-K\"ahler Calabi Metrics, $L^2$ Harmonic Forms, \\
Resolved M2-branes, and AdS$_4$/CFT$_3$ Correspondence}}

\vspace{10pt}
\auth

\vspace{5pt}
{\hoch{\dagger}\upenn}

\vspace{5pt}
{\hoch{\sharp}\damtp}

%%\vspace{5pt}
%%{\hoch{\dagger} \it Department of Physics and Astronomy, Rutgers University,
%%Piscataway, NJ 08855}

\vspace{5pt}
{\hoch{\star}\umich}

\vspace{5pt}
{\hoch{\ddagger}\tamphys}

\vspace{5pt}
{\hoch{\dagger,\sharp,\ddagger}\ihp}

\vspace{10pt}

\underline{ABSTRACT}
\end{center}

     We obtain a simple explicit expression for the hyper-K\"ahler
Calabi metric on the co-tangent bundle of $\CP^{n+1}$, for all $n$, in
which it is constructed as a metric of cohomogeneity one with
$SU(n+2)/U(n)$ principal orbits.  These results enable us to obtain
explicit expressions for an $L^2$-normalisable harmonic 4-form in
$D=8$, and an $L^2$-normalisable harmonic 6-form in $D=12$.  We use
the former in order to obtain an explicit resolved M2-brane solution,
and we show that this solution is invariant under all three of the
supersymmetries associated with the covariantly-constant spinors in
the 8-dimensional Calabi metric.  We give some discussion of the
corresponding dual ${\cal N}=3$ three-dimensional field theory.
Various other topics are also addressed, including superpotentials for
the Calabi metrics and the metrics of exceptional $G_2$ and Spin(7)
holonomy in $D=7$ and $D=8$.  We also present complex and quaternionic
conifold constructions, associated with the cone metrics whose
resolutions are provided by the Stenzel $T^*S^{n+1}$ and Calabi
$T^*\CP^{n+1}$ metrics.  In the latter case we relate the construction
to the hyper-K\"ahler quotient.  We then use the hyper-K\"ahler
quotient to give a quaternionic rederivation of the Calabi metrics.

%{\vfill\leftline{}\vfill
%\vskip 5pt
%\footnoterule
%{\footnotesize \hoch{1} Research supported in part by DOE grant
%DE-FG02-95ER40893 and NATO grant 976951. \vskip -12pt} \vskip 14pt
%{\footnotesize \hoch{2} Research supported in full by DOE grant
%DE-FG02-95ER40899 \vskip -12pt} \vskip 14pt
%{\footnotesize  \hoch{3} Research supported in part by DOE
%grant DE-FG03-95ER40917.\vskip  -12pt}}

%\baselineskip=24pt
\pagebreak
\setcounter{page}{1}

\tableofcontents
\vfill\eject

\section{Introduction}

    The conjectured AdS/CFT correspondence \cite{malda,gkp,wit} asserts
that a background of the form AdS$_d\times {\cal M}_m$ in string theory
or M-theory will have an associated dual description as a conformal
field theory on the $(d-1)$-dimensional boundary of AdS$_d$.  Work in
this area has focused predominantly on the case where the
$m$-dimensional Einstein space ${\cal M}_m$ is a sphere, implying that
the supersymmetry of the CFT will be maximised.  However, if ${\cal
M}_m$ is taken to be some other Einstein space that admits Killing
spinors, then the associated dual CFT will again be supersymmetric, but
with a lesser degree of supersymmetry.  Any such background can
be viewed as the decoupling limit of a $(d-1)$-brane solution, in which
the usual flat space transverse to the brane is replaced by the
Ricci-flat metric $d\hat s^2$ on the cone over ${\cal M}_m$,
%%%%%
\be
d\hat s^2 = dr^2 + r^2\, d\Sigma^2\,,
\ee
%%%%%
where $d\Sigma^2$ is the Einstein metric on ${\cal M}_m$ with
Ricci-tensor $R_{ab} = (m-1)\, g_{ab}$.  This was discussed in detail
for type IIB backgrounds of the form AdS$_5\times {\cal M}_5$, viewed as
the decoupling limit of D3-branes, with ${\cal M}_5$ taken to be the
$U(1)$ bundle over $S^2\times S^2$ known as $T^{1,1}$, in
\cite{klewit}.

    Ricci-flat cone metrics are singular at the apex of the cone, where
the ${\cal M}_m$ principal orbits degenerate to a point.  In many
cases a ``resolution'' of the singularity is possible, in which the
degeneration near the origin is instead locally of the form 
%%%%%
\be
{\cal M}_m \longrightarrow S^p\times \wtd{\cal M}_q\,,\qquad p+q=m\,,
\label{degen1}
\ee
%%%%%
where $S^p$ denotes a round $p$-sphere whose radius tends to zero at the
origin, and $\wtd{\cal M}_q$ is a manifold whose size remains
non-vanishing at the origin.  Provided that the rate of collapse of the
$S^p$ factor is appropriate, the region near the origin smoothly
approaches $\R^{p+1}\times \wtd{\cal M}_q$ locally.\footnote{There is,
of course, always a ``trivial'' cone metric over the round sphere,
${\cal M}_m=S^m$, for which the ``conifold'' is just the
$(m+1)$-dimensional Euclidean manifold, and there is no singularity at
the apex.  This corresponds to the special case $q=0$.}  At large
distance the resolved metric asymptotically approaches the original cone
metric.  Since ${\cal M}_m$ is usually a homogeneous manifold in the
cases of interest (where ${\cal M}_m$ admits Killing spinors), it
follows that the resolutions of the cone metrics will then be described by
manifolds of cohomogeneity one, with a compact symmetry group acting
transitively on the principal orbits.

   In the case of ${\cal M}_5=T^{1,1}$, complete resolved Ricci-flat
K\"ahler metrics are known for which the degeneration (\ref{degen1})
takes any of the forms
%%%%%
\be
T^{1,1}\longrightarrow S^2\times S^3\,,\qquad S^3\times S^2 \,,\qquad
S^1\times (S^2\times S^2)\,.
\ee
%%%%%
A further possibility that now arises is to consider additional branes
in the type IIB theory, that can wrap around the non-trivial cycle in
the transverse space.  This gives rise to so-called ``fractional
D3-branes'' \cite{klebtsey,klebstra}.  In terms of the description
within type IIB supergravity, the wrapping of additional branes
corresponds to turning on additional fluxes for the R-R and NS-NS
3-forms, which are now set equal to the real and imaginary parts of a
self-dual harmonic 3-form in the resolved transverse-space metric.
Physically, one of the consequences of turning on the additional flux is
to perturb the background away from AdS$_5\times T^{1,1}$ in the
decoupling limit, meaning that the dual field theory on the boundary is
perturbed away from a conformally-invariant one.  This has the important
consequence that scale invariance is broken, and a mass gap can now open
up in the spectrum of the operator determining the glueball mass.  Thus
the non-conformal phase with the additional flux is a confining phase.

    Another situation of considerable interest is M-theory backgrounds
of the form AdS$_4\times {\cal M}_7$, which correspond to decoupling
limits of M2-branes.  A detailed discussion of the conifolds where
${\cal M}_7$ is taken to be $Q^{1,1,1}$ (a $U(1)$ bundle over $S^2\times
S^2\times S^2$ with winding numbers 1 over each $S^2$) or $M^{2,3}$ (a
$U(1)$ bundle over $S^2\times \CP^2$ with winding numbers 2 and 3 over
$S^2$ and $\CP^2$) was given in \cite{ffgrtzz}.  For these, and various
other choices for the Einstein manifold ${\cal M}_7$, one can again
replace the singular cone metrics on the 8-dimensional transverse spaces
by smoothed-out Ricci-flat resolutions.  There are three basic kinds of
such irreducible 8-manifolds that admit covariantly-constant spinors,
namely those with exceptional Spin(7) holonomy, Ricci-flat K\"ahler
holonomy $SU(4)$, and hyper-K\"ahler holonomy $Sp(2)\equiv$ Spin(5).
These admit ${\cal N}=1$, 2 and 3 covariantly-constant spinors
respectively.

    Resolved M2-brane solutions using transverse 8-spaces that are
resolutions with Spin(7) and $SU(4)$ holonomy were constructed in
\cite{clpres,cglp1,cglp2}.  In the Spin(7) example the metric is a
resolution of a cone over $S^7$.  However, the metric is not just the
trivial Euclidean one because the Einstein metric on the 7-sphere at the
base of the cone is the non-standard ``squashed'' one, and not the usual
round-sphere metric.  Ricci-flat K\"ahler examples with $SU(4)$ holonomy
were constructed for cases where ${\cal M}_7$ is $Q^{1,1,1}$ or
$M^{2,3}$.  In another example ${\cal M}_7$ was again locally $S^7$, but
this time although the base of the cone has the round $S^7$ metric, its
global topology is actually $S^7/\Z_4$ and so again the resolved
8-manifold is not merely the trivial Euclidean space.\footnote{In fact
this example is a higher-dimensional analogue of the 4-dimensional
Eguchi-Hanson metric \cite{egha}, for which the principal orbits were
shown in \cite{begipapo} to be $S^3/\Z_2$ rather than $S^3$.}  Another
example of a similar kind was based on the manifold ${\cal
M}_7=SU(3)/U(1)$.  In all these cases, there exist
$L^2$-normalisable harmonic 4-forms in resolved transverse 8-manifolds,
and these can be used in order to modify the usual 4-form field strength
background of an M2-brane, leading to deformed M2-brane solutions.  The
notion of ``wrapping'' does not arise here, and correspondingly the
additional field-strength contribution gives no flux integral at
infinity.  The physical interpretation of these deformed M2-brane
solutions is therefore rather different from that for the fractional
D3-branes.

   The principal theme in the present paper is to study the third type
of resolved M2-brane solution mentioned above, where the transverse
8-manifold is hype-K\"ahler, with Spin(5) holonomy.  In fact it can be
shown that there is only one such complete non-singular 
8-metric of cohomogeneity one
\cite{danswa}, namely the hyper-K\"ahler Calabi metric \cite{calabi}.
In order to give an explicit construction of the associated deformed
M2-brane solution it is necessary to find an explicit expression for
the harmonic 4-form.  The usual presentations of the Calabi metric do
not lend themselves to obtaining such explicit results in any
convenient way.  Accordingly, we begin the paper by finding a more
explicit way to construct the Calabi metric.  In fact we shall give a
general construction for the hyper-K\"ahler Calabi metrics in all
dimensions $D=4n+4$.  The manifolds for these solutions are
$T^*\CP^{n+1}$, the co-tangent bundles of the complex projective
spaces $\CP^{n+1}$.  Our approach is along the lines of the
construction in \cite{GP,danswa}.  In an earlier paper \cite{cglp1} we
applied these methods in order to produce fully explicit expressions
for the Stenzel metrics on $T^* S^{n+1}$ \cite{sten}, for which the
principal orbits were $SO(n+2)/SO(n)$.  In this paper we apply a
similar technique to the case of $T^*\CP^{n+1}$, for which the
principal orbits are $SU(n+2)/U(n)$.  

   Our procedure for obtaining the hyper-K\"ahler solutions is by
making an ansatz for metrics of cohomogeneity one whose principal
orbits are $SU(n+2)/U(n)$, with undetermined functions of $r$
parameterising the various radii in the level sets.  Following the
procedure of \cite{danswa}, when obtain first-order equations for
these functions from the integrability conditions for the existence of 
a hyper-K\"ahler structure.  We find that the solution of these
equations leads to the following simple expression for the
hyper-K\"ahler Calabi metric in $D=4n+4$ dimensions:
%%%%%
\crampest
\be 
ds^2 = \fft{dr^2}{1-r^{-4}} + \ft14 r^2\, (1-r^{-4})\, \lambda^2 + 
   r^2\, (\nu_1^2+\nu_2^2) + 
    \ft12 (r^2-1)\, (\sigma_{1\a}^2 + \sigma_{2\a}^2) +
    \ft12 (r^2+1)\, (\Sigma_{1\a}^2 + \Sigma_{2\a}^2) \,,
\label{calabimetric}
\ee
\uncramp
%%%%%
where $(\lambda,\nu_1,\nu_2,\sigma_{1\a},\sigma_{2\a},
\Sigma_{1\a},\Sigma_{2\a})$, with $\a$ running over $n$ values, are
left-invariant 1-forms on the coset $SU(n+2)/U(n)$.  

   We also show how the Einstein equations for the metric ansatz,
viewed as a Lagrangian system, can be cast into a form where they can
be derived from a superpotential, whose associated first-order
equations again admit the hyper-K\"ahler Calabi metrics as solutions.  
In fact by making restrictions in the original parameterisation of the
metrics we can obtain two inequivalent superpotentials, one leading to
the hyper-K\"ahler solutions, and the other to Ricci-flat K\"ahler
solutions on the complex line bundles over $SU(n+2)/(U(n)\times
U(1))$.  

    Having obtained explicit expressions for the hyper-K\"ahler Calabi
metrics in a fashion that is suitable for our purposes, we then look for
middle-dimension harmonic forms.  In particular, we obtain explicit
results for an $L^2$-normalisable (anti) self-dual harmonic 4-form in
the $D=8$ metric, and an $L^2$-normalisable (anti) self-dual harmonic
6-form in the $D=12$ metric.  The $D=8$ result allows us to build a
resolved M2-brane solution.  We also show that all three of the
covariantly-constant spinors of the hyper-K\"ahler 8-metric remain
Killing spinors of the resolved M2-brane solution.  This implies that
the dual 3-dimensional gauge theory on the boundary of AdS$_4$ will have
${\cal N}=3$ supersymmetry.

    In the remainder of the paper, we give some discussion of
the physical significance of this, and previously-obtained resolved
brane solutions.  We analyse the symmetry groups of the dual field
theories.  In particular, we emphasise that one of the purposes of the
brane resolution is to break the conformal symmetry in order to achieve 
confinement.  This leads to very different mechanisms for confinement in
$D=4$ and $D=3$ gauge theories.  In $D=4$, an additional fractional
5-brane flux is needed for the brane-resolution, whilst in $D=3$ the
breaking of the conformal phase is caused by a perturbation of
relevant operators, associated with the pseudoscalar fields of the gauge
theory in the Higgs branch.

   After a concluding section the paper ends with a set of appendices.
In appendix A we present a summary of complete non-compact Ricci-flat
metrics of cohomogeneity one in dimensions $D=4$, 6, 7 and 8.  These are
the dimensions of greatest interest from the point of view of
constructing brane resolutions in string theory and M-theory.  We
include derivations of superpotential formulations for the Einstein
equations leading to metrics of $G_2$ holonomy in $D=7$, and Spin(7)
holonomy in $D=8$.  Appendix B contains a summary of some results for
higher-dimensional generalisations of the Taub-NUT metric.  Appendix C
contains some further results for the $T^*\CP^{n+1}$ construction in the
specific case of $n=1$, where more general superpotentials can be found.
Finally, in appendix D we present a construction of complex and
quaternionic conifolds.  In particular, we show how the Einstein metric
on the base of the quaternionic cone, which is the asymptotic form of
the principal orbits in the Calabi metrics, arises as the hyper-K\"ahler
quotient of the flat metric on $\H^{\, n+2}$.  We then use the
hyper-K\"ahler quotient construction to give a quaternionic
rederivation of the Calabi metric in the form (\ref{calabimetric}).

\section{Metrics on co-tangent bundle of $\CP^{n+1}$}

   In this section, we present an explicit construction of metrics of
cohomogeneity one on the co-tangent bundle of $\CP^n$.  Following the
approach of \cite{danswa}, we then show how the conditions for
Ricci-flatness can be reduced to a system of first-order equations, by
requiring the existence of a hyper-K\"ahler structure.  We obtain
solutions to these equations, and thereby obtain fully explicit
expressions for the hyper-K\"ahler Calabi metrics that will be used in
subsequent sections.  (Later, in appendix D, we shall rederive the
same metrics using the hyper-K\"ahler quotient construction.)

\subsection{Metrics of cohomogeneity one on $T^*\CP^{n+1}$}

    Paralleling the construction of the Stenzel metrics on
$T^*(S^{n+1})$ in \cite{cglp1}, 
here we start with the generators of
$SU(n+2)$, and their associated left-invariant 1-forms $L_A{}^B$,
where $L_A{}^A=0$ and $(L_A{}^B)^\dagger=L_B{}^A$, which satisfy the
exterior algebra
%%%%%
\be
 d L_A{}^B = \im\, L_A{}^C\wedge L_C{}^B\,.\label{sun2alg}
\ee
%%%%%
We then split the $SU(n+2)$ index $A$ as $A=(1,2,\a)$.  The idea will
be to define generators in the coset $SU(n+2)/U(n)$, and those that
lie in the denominator group $U(n)$.  Clearly the latter will include 
the $SU(n)$ generators 
%%%%%
\be
\wtd L_\a{}^\beta \equiv L_\a{}^\beta + \fft1{n}\, Q\,
\delta_\a^\beta\,,\label{sun}
\ee
%%%%%
where the trace subtraction is written in terms of the $U(1)$
generator $Q$, defined to be
%%%%%
\be
Q\equiv L_1{}^1 + L_2{}^2\,.
\ee
%%%%%
There is one other $U(1)$ generator, namely
%%%%%
\be
\lambda\equiv L_1{}^1 - L_2{}^2\,.
\ee
%%%%%
The generators of the coset will then be the complement of $\wtd
L_\a{}^\beta$ and $Q$, namely $\lambda$ and
%%%%%
\be
\sigma^\a \equiv L_1{}^\a\,,\qquad \Sigma^\a \equiv L_1{}^\a\,, \qquad
\nu\equiv L_1{}^2\,.\label{coset}
\ee
%%%%%
Note that $\sigma^\a$, $\Sigma^\a$ and $\nu$ are all complex, while
$\lambda$ is real.

    In general, we find on decomposing (\ref{sun2alg}) that the
exterior derivatives are given by
%%%%%
\bea
d\sigma^\a &=& \ft{\im}{2}\, \lambda\wedge \sigma^\a + \im\, \nu\wedge
\Sigma^\a + \ft{\im}{2} (1+2/n)\, Q\wedge \sigma^\a +
\im\, \sigma^\beta\wedge \wtd L_\beta{}^\a \,,\nn\\
d\Sigma^\a &=& -\ft{\im}{2}\, \lambda\wedge \Sigma^\a + \im\, \bar\nu\wedge
\sigma^\a + \ft{\im}{2} (1+2/n)\, Q\wedge \Sigma^\a +
\im\,\Sigma^\beta\wedge \wtd L_\beta{}^\a \,,\nn\\
d\nu &=& \im\, \lambda\wedge \nu + \im\, \sigma^\a\wedge
\bar\Sigma_\a\,,\nn\\
d\lambda &=& 2\im\, \nu\wedge \bar\nu + \im\, \sigma^\a \wedge
\bar\sigma_\a - \im\, \Sigma^\a\wedge \bar\Sigma_\a\,,\label{decomposed}\\
dQ &=& \im\, \sigma^\a\wedge \bar\sigma_\a + \im\, \Sigma^\a\wedge
\bar \Sigma_\a\,,\nn\\
d\wtd L_\a{}^\beta &=& \im\, \bar\sigma_\a\wedge \sigma^\beta +
   \im\, \bar\Sigma_\a\wedge \Sigma^\beta - \fft1{n}\, 
\im\, (\bar\sigma_\gamma\wedge \sigma^\gamma +
   \bar\Sigma_\gamma\wedge \Sigma^\gamma)\, \delta_\a^\beta  
    +\im\, \wtd L_\a{}^\gamma\wedge \wtd L_\gamma{}^\beta\,.\nn
\eea
%%%%%
 
    The $(4n+4)$-dimensional metric of cohomogeneity one 
on the co-tangent bundle of $\CP^{n+1}$ will be written as
%%%%%
\be ds^2 = dt^2 + a^2\, |\sigma^\a|^2 + b^2\, |\Sigma^\a|^2 + c^2\,
|\nu|^2 + f^2\, \lambda^2\,, \label{genmet}
\ee
%%%%%
where the metric functions $a$, $b$, $c$ and $f$ all depend only on
$t$.  

\subsection{Curvature calculations}

   Define real 1-forms as follows:
%%%%%
\be
\sigma_\a = \sigma_{1\a} + \im\, \sigma_{2\a}\,,\qquad 
\Sigma_\a = \Sigma_{1\a} + \im\, \Sigma_{2\a}\,,\qquad 
\nu=\nu_1 + \im\, \nu_2\,,
\ee
%%%%%
which therefore satisfy 
%%%%%
\bea
d\sigma_{1\a} &=& -\ft12 \lambda\wedge \sigma_{2\a} 
        - \nu_1\wedge \Sigma_{2\a} -
\nu_2\wedge \Sigma_{1\a} - \hat Q\wedge \sigma_{2\a}
-\sigma_{1\beta}\wedge L_{2\beta}{}^\a - \sigma_{2\beta}\wedge 
L_{1\beta}{}^\a \,,\nn\\
d\sigma_{2\a} &=& \ft12 \lambda\wedge \sigma_{1\a} 
        + \nu_1\wedge \Sigma_{1\a} -
\nu_2\wedge \Sigma_{2\a} + \hat Q\wedge \sigma_{2\a}
+\sigma_{1\beta}\wedge L_{1\beta}{}^\a - \sigma_{2\beta}\wedge 
L_{2\beta}{}^\a \,,\nn\\
d\Sigma_{1\a} &=& \ft12 \lambda\wedge \Sigma_{2\a} - \nu_1\wedge 
\sigma_{2\a} + \nu_2\wedge \sigma_{1\a} -\hat Q\wedge \Sigma_{2\a}
-\Sigma_{1\beta}\wedge L_{2\beta}{}^\a - 
\Sigma_{2\beta}\wedge L_{1\beta}{}^\a\,,\nn\\
d\Sigma_{2\a} &=& -\ft12 \lambda\wedge \Sigma_{1\a} + \nu_1\wedge 
\sigma_{1\a} + \nu_2\wedge \sigma_{2\a} +\hat Q\wedge \Sigma_{1\a}
+\Sigma_{1\beta}\wedge L_{1\beta}{}^\a - 
\Sigma_{2\beta}\wedge L_{2\beta}{}^\a\,,\nn\\
d\nu_1 &=& -\lambda\wedge \nu_2 + \sigma_{1\a}\wedge \Sigma_{2\a} -
\sigma_{2\a}\wedge \Sigma_{1\a}\,,\label{genext1}\\
d\nu_2 &=& \lambda\wedge \nu_1 + \sigma_{1\a}\wedge \Sigma_{1\a} +
\sigma_{2\a}\wedge \Sigma_{2\a}\,,\nn\\
d\lambda &=& 2\sigma_{1\a}\wedge \sigma_{2\a} - 
       2 \Sigma_{1\a}\wedge \Sigma_{2\a}  
    + 4 \nu_1\wedge \nu_2\,,\nn
\eea
%%%%%
where we have defined $\hat Q = \ft12(1+2/n)\, Q$, and $\wtd L_\a{}^\beta =
L_{1\a}^\beta + \im\, L_{2\a}{}^\beta$.  Note that since $\wtd L_\a{}^\beta$
is hermitean, it follows that $L_{1\a}{}^\beta$ is symmetric, and
$L_{2\a}{}^\beta$ is antisymmetric.  

    In addition, we need to know that for $\hat Q$, $L_{1\a}{}^\beta$ and 
$L_{2\a}{}^\beta$, which lie outside the coset, we have
%%%%%
\bea
d\hat Q &=&(1+\fft2{n})\, ( \sigma_{1\a}\wedge \sigma_{2\a} 
+ \Sigma_{1\a}\wedge\Sigma_{2\a})\,,\nn\\
dL_{1\a}{}^\beta &=& -\sigma_{1\a}\wedge\sigma_{2\beta}
-\sigma_{1\beta}\wedge \sigma_{2\a} -\Sigma_{1\a}\wedge\Sigma_{2\beta}
-\Sigma_{1\beta}\wedge \Sigma_{2\a} \nn\\
&&+\fft2{n}\, 
(\sigma_{1\gamma}\wedge\sigma_{2\gamma}+
      \Sigma_{1\gamma}\wedge \Sigma_{2\gamma})\, \delta_\a{}^\beta 
      -L_{1\a}{}^\gamma\wedge L_{2\gamma}{}^\beta -
        L_{2\a}{}^\gamma\wedge L_{1\gamma}{}^\beta\,,\label{genext2}\\
dL_{2\a}{}^\beta &=& \sigma_{1\a}\wedge\sigma_{1\beta}
+\sigma_{2\a}\wedge \sigma_{2\beta} +\Sigma_{1\a}\wedge\Sigma_{1\beta}
+\Sigma_{2\a}\wedge \Sigma_{2\beta} +L_{1\a}{}^\gamma\wedge 
L_{1\gamma}{}^\beta - L_{2\a}{}^\gamma\wedge L_{2\gamma}{}^\beta
\,.\nn
\eea
%%%%%

   In terms of the real 1-forms, the metric (\ref{genmet}) becomes
%%%%%
\be
ds^2 = dt^2 + a^2\, (\sigma_{1\a}^2 + \sigma_{2\a}^2) +
b^2\, (\Sigma_{1\a}^2 + \Sigma_{2\a}^2) +
c^2\, (\nu_1^2 + \nu_2^2) + f^2\, \lambda^2\,.\label{genmetn}
\ee
%%%%%
In the obvious orthonormal basis, 
%%%%%
\bea
&&e^0=dt\,,\quad e^{\td 0} = f\, \lambda\,,\quad e^{\td 1}=
c\, \nu_1\,,\quad e^{\td 2}=c\, \nu_2\,,\nn\\
&&e^{1\a}=a\, \sigma_{1\a}\,,\quad e^{2\a}=a\, \sigma_{2\a}\,,\quad 
e^{1\td\a}=b\, \Sigma_{1\a}\,,\quad e^{2\td\a}=b\, \Sigma_{2\a}\,,
\label{genorth}
\eea
%%%%%
we find after some calculation that
the Ricci tensor is diagonal, and there are just five inequivalent
eigenvalues, associated with the $dt$, $\sigma_{i\a}$, $\Sigma_{i\a}$,
$\nu_i$ and $\lambda$ directions respectively.  In an obvious
notation, these are
%%%%%
\bea
R_0 &=& -2n\, \Big(\fft{\ddot a}{a} + \fft{\ddot b}{b}\Big) -
\fft{2\ddot c}{c} - \fft{\ddot f}{f}\,,\nn\\
R_a &=& -\fft{\ddot a}{a} - \fft{(2n-1)\,{\dot a}^2}{a^2} -
\fft{2n\, \dot a\, \dot b}{a\, b} - \fft{2\dot a\, \dot c}{a\, c} -
\fft{\dot a\, \dot f}{a\, f} + \fft{a^4-b^4-c^4}{a^2\, b^2\, c^2} + 
  \fft{2(n+2)}{a^2} - \fft{2 f^2}{a^4}\,,\nn\\
R_b &=& -\fft{\ddot b}{b} - \fft{(2n-1)\, {\dot b}^2}{b^2} -
\fft{2n\, \dot a\, \dot b}{a\, b} - \fft{2\dot b\, \dot c}{b\, c} -
\fft{\dot b\, \dot f}{b\, f} + \fft{b^4-a^4-c^4}{a^2\, b^2\, c^2} + 
  \fft{2(n+2)}{b^2} - \fft{2 f^2}{b^4}\,,\nn\\
R_c &=& -\fft{\ddot c}{c} - \fft{{\dot c}^2}{c^2} -
\fft{2n\, \dot a\, \dot c}{a\, c} -  \fft{2n\, \dot b\, \dot c}{b\, c} -
\fft{\dot c\, \dot f}{c\, f} +
   n\,\Big( \fft{c^4-a^4-b^4}{a^2\, b^2\, c^2}\Big) + 
  \fft{2(n+2)}{c^2} - \fft{8 f^2}{c^4}\,,\nn\\
R_f &=& -\fft{\ddot f}{f} - \fft{2n\, \dot a\, \dot f}{a\, f} -
\fft{2n\, \dot b\, \dot f}{b\, f} - \fft{2\dot c\, \dot f}{c\, f} 
+ f^2\, \Big( \fft{2n}{a^4} + \fft{2n}{b^4} + \fft{8}{c^4}\Big)\,.
\label{riccigen}
\eea
%%%%%

\subsection{Hyper-K\"ahler solutions of the Einstein equations}

   We are principally interested in obtaining the hyper-K\"ahler Calabi
metrics, which should arise as particular solutions of the Ricci-flat
Einstein equations obtained above.  The easiest way to find these
metrics is to follow the procedure of Dancer and Swann \cite{danswa}, of
writing down the ans\"atze for the three hyper-K\"ahler forms, and then
imposing the conditions of covariant-constancy.

   We seek three simultaneous K\"ahler forms $J_i$. 
It quickly becomes evident, after consulting (\ref{genext1}), that the
following are three candidates for the K\"ahler forms:
%%%%%
\bea
J_1 &=& f\, dt\wedge\lambda +c^2\, \nu_1\wedge \nu_2
+ a^2 \, \sigma_{1\a}\wedge \sigma_{2\a}
    -b^2\, \Sigma_{1\a}\wedge \Sigma_{2\a} \,,\nn\\
J_2 &=& c\, dt\wedge\nu_1 -c\, f\, \lambda\wedge \nu_2
+ a\, b \, (\sigma_{1\a}\wedge \Sigma_{2\a}
   -  \sigma_{2\a}\wedge \Sigma_{1\a}) \,,\label{3j}\\
J_3 &=& c\, dt\wedge\nu_2  + c\, f\, \lambda\wedge \nu_1
+ a\, b \, (\sigma_{1\a}\wedge \Sigma_{1\a}
    + \sigma_{2\a}\wedge \Sigma_{2\a}) \,.\nn
\eea
%%%%%
Note that the $a$, $b$, $c$ and $f$ prefactors are determined
completely, up to signs, by the fact that each term in
each $J_i$ should be just a wedge product of vielbeins (see
(\ref{genorth}).  The structure of the $dt$, $\lambda$, $\nu_1$ and
$\nu_2$ terms in each case is determined by the requirement that these
must supply the ``remainder'' of the K\"ahler form, which must span
the total dimension of the manifold, and so structurally these terms
follow once the terms involving $\sigma_{i\a}$ and $\Sigma_{i\a}$ are settled.
The precise details can be fixed easily, by looking
first at the terms involving $dt$, when one imposes $dJ_i=0$.  Note
also that we have already incorporated some of the easily-determined
results from these equations, in making the specific selections of
$\pm$ signs presented in (\ref{3j}).

    Now, imposing $dJ_i=0$, we find that $dJ_1=0$ leads to the equations
%%%%%
\be
\fft{d\, (a^2)}{dt} = 2f\,,\qquad 
\fft{d\, (b^2)}{dt} = 2f\,,\qquad 
\fft{d\, (c^2)}{dt} = 4f\,,\qquad 
a^2+b^2 = c^2\,.\label{dseq1}
\ee
%%%%%
Similarly, we find that imposing $dJ_2=0$ gives
%%%%%
\be
\fft{d\, (a\, b)}{dt} = c\,,\qquad 
\fft{d\, (c\, f)}{dt} = c\,,\qquad 
a\, b = c\, f\,.\label{dseq2}
\ee
%%%%%
Finally, imposing $dJ_3=0$ just gives the same equations as from $dJ_2=0$.
We see that (\ref{dseq1}) and (\ref{dseq2}) are precisely the
first-order equations obtained by Dancer and Swann \cite{danswa}.

     It is straightforward to solve these equations.  In terms of the 
new radial variable $r$, related to $t$ by $dt=h\, dr$, we find 
%%%%%
\be
a^2=\ft12(r^2-1)\,,\quad b^2=\ft12 (r^2+1)\,,\quad c^2=r^2\,,\quad 
f^2= \ft14 r^2\,(1-r^{-4})\,,\quad h^2 = (1-r^{-4})^{-1}\,.\label{calsol}
\ee
%%%%%
The Calabi metric in $D=4n+4$ dimensions is then given by
(\ref{genmetn}). One can easily verify that the Ricci tensor, given by
(\ref{riccigen}), does indeed vanish.  Note that, remarkably, the
expressions for the metric functions are completely independent of the
dimension $D=4n+4$.  In fact by specialising to $n=0$, in which case
there are no 1-forms $\sigma^\a$ or $\Sigma^\a$ at all, we recover the
well-known Eguchi-Hanson metric.  The characteristic $(1-r^{-4})$
functions of this metric thus continue to appear in all the
higher-dimensional hyper-K\"ahler Calabi metrics.

\subsection{Geometry of the Calabi metrics}

   It is evident from (\ref{genmetn}) and (\ref{calsol}) that the radial
coordinate runs from $r=1$ to $r=\infty$.  Asymptotically, at large $r$,
the metric approaches the cone over the homogeneous tri-Sasaki 
Einstein manifold $SU(n+2)/U(n)$:
%%%%%
\be
ds^2 \sim dr^2 + \ft14 r^2\,( \lambda^2 
+ 2 |\sigma_\a|^2 +2 |\Sigma_\a|^2 + 4 |\nu|^2) \,.
\ee
%%%%%

   Near to $r=1$, it is helpful to introduce a new radial coordinate
$\rho$, defined by $\ft12(r^2-1) = \rho^2$.  In terms of this, we find
that near $\rho=0$ the metric approaches
%%%%%
\be
ds^2 \sim d\rho^2 + \rho^2\, \lambda^2 + \rho^2 \, |\sigma_\a|^2 +
|\Sigma_\a|^2 +  |\nu|^2 \,,
\ee
%%%%%
and so the principal $SU(n+2)/U(n)$ orbits collapse down 
to a $\CP^{n+1}$ bolt.  The geometry near $\rho=0$ is locally
$\R^{2n+2}\times \CP^{n+1}$.   

\subsection{Complex structures in the Calabi metrics}

    The Calabi metrics are hyper-K\"ahler, and so there are three
complex structure tensors, whose multiplication rules are those of the
imaginary unit quaternions.  After lowering the upper index in each
case to give a 2-form, we see from (\ref{genorth}) and (\ref{3j}) that
the three associated K\"ahler forms are
%%%%%
\bea
J_1 &=& e^0\wedge e^{\td 0} + e^{\td 1}\wedge e^{\td 2} +
e^{1\a}\wedge e^{2\a} - e^{1\td\a}\wedge e^{2\td\a}\,,\nn\\
J_2 &=& e^0\wedge e^{\td 1} - e^{\td 0}\wedge e^{\td 2} +
e^{1\a}\wedge e^{2\td\a} - e^{2\a}\wedge e^{1\td\a}\,,\label{3k}\\
J_3 &=& e^0\wedge e^{\td 2} + e^{\td 0}\wedge e^{\td 1} +
e^{1\a}\wedge e^{1\td\a} + e^{2\a}\wedge e^{2\td\a}\,.\nn
\eea
%%%%%

    If we choose $J=J_1$ to define a privileged complex structure, then 
we see from (\ref{3k}) that we can define a complex vielbein basis
%%%%%
\be
\ep^0 \equiv  e^0+\im\, e^{\td 0}\,,\qquad 
\ep^{\#} \equiv  e^1+ \im\, e^2\,,\qquad \ep^\a \equiv e^{1\a} + \im\,
e^{2\a} \,,\qquad \ep^{\td\a} \equiv e^{1\td\a} -\im\, e^{2\td\a}\,,
\label{complexb}
\ee
%%%%%
in terms of which 
%%%%%
\be
J_1 = \ft{\im}{2}\, (\ep^0\wedge \bar\ep^0 + \ep^\#\wedge \bar\ep^\# 
+ \ep^\a\wedge \bar\ep^\a + \ep^{\td\a}\wedge \bar\ep^{\td\a})\,,
\ee
%%%%%
which is, of course, of type $(1,1)$.   The two complex combinations
$K_\pm \equiv J_2\pm\im \, J_3$ of the other two K\"ahler forms are
then given by
%%%%%
\be
K_+ = \ep^0\wedge\ep^\# + \im\, \ep^\a\wedge \ep^{\td\a}\,,\qquad
 K_- = \bar\ep^0\wedge\bar\ep^\# - 
\im\, \bar\ep^\a\wedge \bar\ep^{\td\a}\,.
\ee
%%%%%
These are pure $(2,0)$ and $(0,2)$ respectively, as expected.

     Note that the $SU(n+2)$ factor in the $U(n+2)$ isometry group
acts triholomorphically; \ie it leaves invariant all three complex
structures in (\ref{3k}).  The remaining $U(1)$ factor in the isometry
group leaves $J_1$ invariant, while rotating $J_2$ into $J_3$. 

   Using (\ref{genext1}) and (\ref{genext2}), we can write
the K\"ahler forms $J_i$ locally in terms of 1-form potentials, $J_i=dA_i$,
with
%%%%%
\be
A_1 = \ft14 r^2\, \lambda - \ft14 Q\,,\qquad 
A_2 = \ft12 (r^4-1)^{1/2}\, \nu_1\,,\qquad 
A_3 =  \ft12 (r^4-1)^{1/2}\, \nu_2\,.\label{3ares}
\ee
%%%%%
Note that unlike $A_2$ and $A_3$, we cannot give a general expression
for $A_1$ using quantities intrinsic to the $SU(n+2)/U(n)$ coset
within the framework we are using here.   Of course if we introduced
coordinates for the coset, $Q$ would then be expressible in terms of them. 
In appendix D, we show how $A_1$ acquires an interpretation as the
connection on the $U(1)$ Hopf fibres in a hyper-K\"ahler quotient
construction of the Calabi metrics.

\section{The eight-dimensional case; Calabi metric on $T^*\CP^2$}

    Our principle motivation for studying the hyper-K\"ahler Calabi
metrics is in order to obtain explicit results for the 8-dimensional
example, which will allow us to construct a new explicit resolved
M2-brane solution.  Accordingly, in this section we shall present some
further details of the 8-dimensional example, including in particular
complete expressions for its Riemann curvature.

\subsection{Conventions and curvature}

   The 8-dimensional case corresponds to setting $n=1$ in the general
results of the previous section.  The $\a$ indices now only takes a
single value, namely $\a=3$, and so the traceless ``$SU(1)$ generators''
$\wtd L_\a{}^\beta$ vanish here.  The coset is $SU(3)/U(1)$.  Rather
than following the general notation of section 2, it is more
convenient here to adopt a different labelling of the real 1-forms,
corresponding to the real and imaginary parts of $\sigma_\a$,
$\Sigma_\a$ (with $\a=3$ here in $D=8$) and $\nu$ as follows:
%%%%%
\be
\sigma^3 \equiv \sigma_1 + \im\, \sigma_2\,,\qquad
\Sigma^3 \equiv \Sigma_1 + \im\, \Sigma_2\,,\qquad
\nu \equiv \nu_1+\im\, \nu_2\,.
\ee
%%%%%
Thus the 8-metric will be
%%%%%
\be
ds_8^2 = dt^2 + a^2\, (\sigma_1^2 +\sigma_2^2) + 
b^2\, (\Sigma_1^2 + \Sigma_2^2) + c^2\,(\nu_1^2 + \nu_2^2) 
 + f^2\, \lambda^2\,,\label{d8calmet}
\ee
%%%%%
with the metric functions given by (\ref{calsol}).
We define the obvious orthonormal basis
%%%%%
\be
e^0=dt\,,\quad e^1 = a\, \sigma_1\,\quad e^2 = a\, \sigma_2\,\quad 
e^3 = b\, \Sigma_1\,\quad e^4 = b\, \Sigma_2\,\quad 
e^5 = c\, \nu_1\,\quad e^6 = c\, \nu_2\,\quad 
e^7 = f\, \lambda\,.\label{ortho}
\ee
%%%%%
(Note that for convenience we are now replacing the previous indices
$(\td 0, \td 1, \td 2)$ of the generic case (\ref{ortho}) by $(7,5,6)$
respectively.) 

    It is useful to note, from the general expressions
in section 2.2, that the exterior derivatives of the left-invariant
1-forms in $D=8$ are given by
%%%%%
\bea
d\sigma_1 &=& -\ft12 \lambda\wedge \sigma_2 - \nu_1\wedge \Sigma_2 -
\nu_2\wedge \Sigma_1 - \ft32 Q\wedge \sigma_2\,,\nn\\
d\sigma_2 &=& \ft12 \lambda\wedge \sigma_1 + \nu_1\wedge \Sigma_1 -
\nu_2\wedge \Sigma_2 + \ft32 Q\wedge \sigma_1\,,\nn\\
d\Sigma_1 &=& \ft12 \lambda\wedge \Sigma_2 - \nu_1\wedge \sigma_2 +
\nu_2\wedge \sigma_1 - \ft32 Q\wedge \Sigma_2\,,\nn\\
d\Sigma_2 &=& -\ft12 \lambda\wedge \Sigma_1 + \nu_1\wedge \sigma_1 +
\nu_2\wedge \sigma_2 + \ft32 Q\wedge \Sigma_1\,,\nn\\
d\nu_1 &=& -\lambda\wedge \nu_2 - \sigma_2\wedge \Sigma_1 +
\sigma_1\wedge \Sigma_2\,,\nn\\
d\nu_2 &=& \lambda\wedge \nu_1 + \sigma_1\wedge \Sigma_1 +
\sigma_2\wedge \Sigma_2\,,\nn\\
d\lambda &=& 2\sigma_1\wedge \sigma_2 - 2 \Sigma_1\wedge \Sigma_2  
    + 4 \nu_1\wedge \nu_2\,.\label{d8ext}
\eea
%%%%%
In addition, we need to know that for $Q$, which lies outside the
coset, we have
%%%%%
\be
dQ = 2\sigma_1\wedge \sigma_2 + 2\Sigma_1\wedge\Sigma_2\,.
\ee
%%%%%

   Using the expressions (\ref{3j}) for the three K\"ahler forms, we
find that in the orthonormal basis (\ref{ortho}) they are given by
%%%%%
\bea
J_1 &=& e^0\wedge e^7 + e^1\wedge e^2 - e^3\wedge e^4 + e^5\wedge e^6
\,,\nn\\
J_2 &=& e^0\wedge e^5 + e^6\wedge e^7 + e^1\wedge e^4 - e^2\wedge
e^3\,,\nn\\
J_3 &=& e^0\wedge e^6 - e^5\wedge e^7 + e^1\wedge e^3 + e^2\wedge e^4
\,.\label{d8kahler}
\eea
%%%%%

     It is straightforward to evaluate the curvature 2-forms
$\Theta_{ab}$ for the 8-dimensional Calabi metric. After some algebra,
we find that they are given by
%%%%% 
\bea
\Theta_{01} &=& -\fft1{r^4}\,(-e^0\wedge e^1 + e^2\wedge e^7 + e^3\wedge
e^6 + e^4\wedge e^5)\,,\nn\\
\Theta_{03} &=& -\fft1{r^4}\, (e^0\wedge e^3 + e^1\wedge e^6 - e^2\wedge
e^5 + e^4\wedge e^7)\,,\nn\\
\Theta_{05} &=& -\fft2{r^4}\, (e^0\wedge e^5 - e^6\wedge e^7)\,,\nn\\
\Theta_{07} &=& \fft2{r^4}\, (e^1\wedge e^2 + e^3\wedge e^4) +
\fft4{r^6}\, (e^0\wedge e^7 - e^5\wedge e^6)\,,\nn\\
\Theta_{12} &=& \fft4{r^2}\, (e^1\wedge e^2 + e^3\wedge e^4) +
\fft2{r^4}\, (e^0\wedge e^7 - e^5\wedge e^6)\,,\nn\\
\Theta_{13}&=& -\fft2{r^2} \, (e^1\wedge e^3 - e^2\wedge e^4)\,,\nn\\
\Theta_{15} &=& -\fft1{r^4}\, (e^0\wedge e^4 + e^1\wedge e^5 +
e^2\wedge e^6 - e^3\wedge e^7)\,,\nn\\
\Theta_{17} &=& \fft1{r^4}\, (e^0\wedge e^2 + e^1\wedge e^7 +
e^3\wedge e^5 - e^4\wedge e^6)\,,\nn\\
\Theta_{34} &=& \fft4{r^2}\, (e^1\wedge e^2 + e^3\wedge e^4) +
\fft2{r^4}\, (e^0\wedge e^7 - e^5\wedge e^6)\,,\nn\\
\Theta_{35} &=& \fft1{r^4}\, (e^0\wedge e^2 + e^1\wedge e^7 +
e^3\wedge e^5 - e^4\wedge e^6)\,,\nn\\
\Theta_{37} &=& \fft1{r^4}\, (e^0\wedge e^4 + e^1\wedge e^5 +
e^2\wedge e^6 - e^3\wedge e^7)\,,\nn\\
\Theta_{56} &=& -\fft2{r^4}\, (e^1\wedge e^2 + e^3\wedge e^4) -
\fft4{r^6}\, (e^0\wedge e^7 - e^5\wedge e^6)\,,\nn\\
\Theta_{57} &=& -\fft2{r^6}\, (e^0\wedge e^6 - e^5\wedge e^7)\,,\label{8curv}
\eea
%%%%%
where we are using the orthonormal basis (\ref{ortho}).  (Note that we
have listed only those which are ``inequivalent.''  For example,
$\Theta_{02}$ will be just like $\Theta_{01}$, with appropriate
renumberings.)  

  It is interesting to note that here, and indeed for the
hyper-K\"ahler Calabi metrics in all dimensions $D=4n+4\ge8$, certain
orthonormal components of the Riemann tensor fall off only as $1/r^2$.
These are components that are absent when $n=0$, and so the curvature
of the Eguchi-Hanson metric falls of exceptionally rapidly, in
comparison to the hyper-K\"ahler Calabi metrics in higher dimensions.

\subsection{Covariantly-constant spinors}

    From the expressions (\ref{8curv}) for the curvature 2-forms of the
8-dimensional Calabi metric, it is easy to study the integrability
conditions $\Theta_{ab}\, \Gamma_{ab}\, \eta=0$ for covariantly-constant
spinors, and hence to determine the number of independent ones.  It is
not hard to establish that in fact the content of these integrability
conditions is completely implied by the following subset of the
conditions, namely
%%%%%
\be
(\Gamma_{07} -\Gamma_{56})\, \eta= (\Gamma_{12}+\Gamma_{34})\, \eta=
(\Gamma_{01} -\Gamma_{27} -\Gamma_{36} -\Gamma_{45})\, \eta=0\,.
\label{gammacon}
\ee
%%%%%
Furthermore, it follows from these that there are exactly 3
linearly-independent covariantly-constant spinors, as one would expect
for a hyper-K\"ahler metric in eight dimensions.  In fact, in the
natural choice of spin frame following from (\ref{ortho}), it follows
that the conditions $D\, \eta \equiv d\, \eta +\ft14 \omega_{ab}\,
\Gamma_{ab}\, \eta=0$ for covariant-constancy reduce, after using
(\ref{gammacon}), to simply $d\, \eta=0$.  Thus in this frame the three
covariantly-constant spinors are just the three solutions of the
algebraic conditions (\ref{gammacon}), with constant components.

\section{Superpotentials and first-order systems}

    In this section, we shall make a more detailed investigation of the
equations resulting from requiring Ricci-flatness for the general class
of metrics of cohomogeneity one on $T^*\CP^{n+1}$ that were introduced
in section 2.  In particular, we shall show how one can obtain the
hyper-K\"ahler Calabi metrics as solutions of a first-order system that
follows from a superpotential.  This is achieved by first making certain
algebraic restrictions on the metric functions $a$, $b$, $c$ and $f$
appearing in the ansatz (\ref{genmet}).  We shall also show how a
different first-order system, with an {\it inequivalent} superpotential,
can be obtained by making a different set of algebraic restrictions on
$a$, $b$, $c$ and $f$.  In this latter case, the first-order system
implies Ricci-flat K\"ahler geometry, with $SU(2n+2)$ holonomy, rather
than the hyper-K\"ahler geometry, with $Sp(n+1)$ holonomy, of the
previous first-order system.  For general dimensions $D=4n+4$ the
unrestricted system, with all four of $a$, $b$, $c$ and $f$ remaining as
independent dynamical variables, appears not to admit a description in
terms of first-order equations and a superpotential.  However, we find
that such a description {\it is} possible for the special case of $n=1$,
corresponding to $D=8$ dimensions.  In fact the resulting first-order
equations imply Spin(7) holonomy, which is an exceptional case in the
Berger classification \cite{berger}.  This presumably accounts for the
non-existence of a superpotential for the general 4-variable system in
dimensions $D=4n+4\ge 12$.  In the following three subsections we derive
the 4-variable superpotential for Spin(7) holonomy in $D=8$, and then in
arbitrary dimensions $D=4n+4$ the 2-variable superpotential for
hyper-K\"ahler $Sp(n+1)$ holonomy and the inequivalent 2-variable
superpotential for Ricci-flat K\"ahler $SU(2n+2)$ holonomy.

\subsection{Superpotential for Spin(7) holonomy in $D=8$}

   By considering the $G_{00}$ component of the Einstein tensor
constructed from $R_{AB}$ in section 2.2, which defines a Hamiltonian
$H=T+V$, we can derive the equations for Ricci-flatness from the
Lagrangian $L=T-V$, together with the constraint $H=0$.  Specialising to
$n=1$, we find that after changing radial variable to $\eta$ defined by
$dt=(a\, b\, c)^2\, f\, d\eta$, $T$ and $V$ are given by
%%%%%
\bea
T &=& 2{\a'}^2 + 2{\beta'}^2 + 2{\gamma'}^2 + 8\a'\, \beta' + 8\a'\,
\gamma' + 8 \beta'\, \gamma' + 4\a'\, \sigma' + 4\beta'\, \sigma' +
4\gamma'\, \sigma' \,,\nn\\
V &=& -12 \Big( \fft{1}{a^2} + \fft{1}{b^2} + \fft{1}{c^2}\Big)
+ 2 \Big(\fft{a^4+b^4+c^4}{a^2\, b^2\, c^2}\Big) + 2 f^2\,
\Big(\fft1{a^4} + \fft1{b^4}\Big) + \fft{8f^2}{c^4}\,,\label{tveq8}
\eea
%%%%%
where $a=e^\a$, $b=e^\beta$, $c=e^\gamma$ and $f=e^\sigma$.  The
prime denotes a derivative with respect to the new radial variable
$\eta$. 

  Defining the ``DeWitt metric'' $g_{ij}$ by $T=\ft12 g_{ij}\,
{\a^i}'\, {\a^j}'$, where $\a^i\equiv(\a,\beta,\gamma,\sigma)$, we
find after some calculation that the potential $V$ can be written in
terms of a superpotential $W$ as 
%%%%%
\be
V= -\ft12 g^{ij}\, \fft{\del W}{\del\a^i}\, \fft{\del W}{\del\a^j}\,,
\label{vfromw}
\ee
%%%%%
with $W$ given by
%%%%%
\be
W= 4 a\, b\, c\, f \, (a^2 + b^2 + c^2 )  - 
2 f^2\, (2 a^2\, b^2 -a^2\, c^2 - b^2\, c^2)\,.\label{spin7spot}
\ee
%%%%%

   The first-order equations following from this superpotential
formulation of the equations, namely ${\a^i}'= g^{ij}\, \del
W/\del\a^j$, are
%%%%%
\bea
\a' &=& -e^{3\a+\beta+\gamma+\sigma} + e^{\a+3\beta+\gamma+\sigma} +
e^{\a+\beta+3\gamma+\sigma} -
 e^{2\beta+2\gamma+2\sigma} \,,\nn\\
\beta' &=& e^{3\a+\beta+\gamma+\sigma} - e^{\a+3\beta+\gamma+\sigma} +
e^{\a+\beta+3\gamma+\sigma} -
 e^{2\a+2\gamma+2\sigma} \,,\nn\\
\gamma' &=& e^{3\a+\beta+\gamma+\sigma} + e^{\a+3\beta+\gamma+\sigma} -
e^{\a+\beta+3\gamma+\sigma} + 2
 e^{2\a+2\beta+2\sigma} \,,\nn\\
\sigma' &=& -2
e^{2\a+2\beta+2\sigma} + e^{2\beta+2\gamma + 2\sigma} + 
    e^{2\a+2\gamma +2\sigma}\,,
\eea
%%%%%
Note that another way of writing them is to go back to the original
radial variable $t$.  In terms of this, we have
%%%%%
\bea
\dot\a &=& \fft{b^2+c^2-a^2}{a\, b\, c} - \fft{f}{a^2}\,,\nn\\
\dot\beta &=& \fft{a^2+c^2-b^2}{a\, b\, c} - \fft{f}{b^2}\,,\nn\\
\dot\gamma &=& \fft{a^2+b^2-c^2}{a\, b\, c} +\fft{2f}{c^2}\,,\nn\\
\dot\sigma &=&-\fft{2f}{c^2}+ \fft{f}{a^2} + \fft{f}{b^2}\,.\label{fo}
\eea
%%%%%

   If we substitute the above first-order equations into the expressions
for the curvature 2-forms, and then calculate the integrability
condition for covariantly-constant spinors, $\Theta_{ab}\, \Gamma_{ab}\,
\eta=0$, we find that there is exactly one solution.  Thus these are the
first-order equations for Spin(7) holonomy.

    It is not clear how to solve these first-order equations in general.
However, a special solution can be obtained by setting $b=a$ and
$f= -c/2$, which, it can be verified, is consistent with (\ref{fo}).  The
equations then reduce to $\dot a= 3c/(2a)$ and $\dot c=1-c^2/a^2$, and
it is easily seen that in terms of a new radial variable $r$ defined by
$dr= c\, dt$, these equations lead to the solution $a^2=9r^2/10$ and
$c^2=1-r^{-10/3}$.  Thus we obtain the 8-dimensional Ricci-flat metric
%%%%%
\be
ds_8^2 = \fft{dr^2}{1-r^{-10/3}} + \fft{9 r^2}{100}\, (1-r^{-10/3})\, 
(\lambda^2 + 4 \nu_1^2 + 4\nu_2^2) +  
\fft{9r^2}{10}\, (\sigma_1^2 +\sigma_2^2+ 
\Sigma_1^2 + \Sigma_2^2)\,.\label{spin7met}
\ee
%%%%%
This is structurally very similar to the 8-metric of Spin(7) holonomy
constructed in \cite{brysal,gibpagpop}, which was defined on the
complete manifold of an $\R^4$ bundle over $S^4$.  The difference here
is that the manifold for (\ref{spin7met}) is the analogous $\R^4$ bundle
over $\CP^2$.  Indeed, one can easily verify by looking at the
integrability conditions $\Theta_{ab}\, \Gamma_{ab}\, \eta=0$ for
covariantly-constant spinors that there is exactly one solution, and so
the metric (\ref{spin7met}) has Spin(7) holonomy.  Unfortunately it
suffers from a conical singularity at $r=1$, since the topology of the
3-dimensional fibres over $\CP^2$, with their metric $(\lambda^2 + 4
\nu_1^2 + 4\nu_2^2)$, is $RP^3$ rather than $S^3$.

\subsection{Lagrangian formulation for the general $D=4n+4$ metrics}

   In dimensions $D=4n+4\ge12$ we can obtain the general 4-variable
Ricci-flat equations from a Lagrangian $L=T-V$, together with the
constraint $T+V=0$, where
%%%%%
\crampest
\bea
T &=& 2n(2n-1)\, ({\a'}^2 + {\beta'}^2) + 2{\gamma'}^2 +
8n^2\, \a'\, \beta' + 8n\, (\a'\, \gamma' + \beta'\,
\gamma') + 4n\, (\a'\, \sigma' + \beta'\, \sigma') +
4\gamma'\, \sigma' \,,\nn\\
V &=& -4(n+2)\, \Big( \fft{n}{a^2} + \fft{n}{b^2} + \fft{1}{c^2}\Big)
+ 2n\, \Big(\fft{a^4+b^4+c^4}{a^2\, b^2\, c^2}\Big) + 2n\, f^2\,
\Big(\fft1{a^4} + \fft1{b^4}\Big) + \fft{8f^2}{c^4}\,.\label{tveq}
\eea
\uncramp
%%%%%
As usual we have written $a=e^\a$, $b=e^\beta$, $c=e^\gamma$,
$f=e^{\sigma}$, and a prime denotes a derivative with respect to the new
radial variable $\eta$ defined by $dt=(a\, b)^{2n}\, c^2\, f\, d\eta$.

   From the kinetic energy in (\ref{tveq}), which may be written as 
$T=\ft12 g_{ij}\, \dot\a^i\, \dot\a^j$, where
$\a^i=(\a,\beta\,\gamma,\sigma)$, we can read off the metric $g_{ij}$;
%%%%%
\be
g_{ij} = \pmatrix{4n(2n-1) & 8n^2 & 8n & 4n\cr
          8n^2 & 4n(2n-1) & 8n & 4n \cr
          8n & 8n & 4 & 4 \cr
          4n & 4n & 4 & 0}\,.\nn\\
\ee
%%%%%

    Although $V$ can be written in terms of a superpotential, as in
(\ref{vfromw}), (\ref{spin7spot}), for the special case $D=4n+4=8$, this
does not appear to be possible for $n\ge2$.  This is not unreasonable,
since the occurrence of Spin(7) as a holonomy group in $D=8$ is
exceptional, with no higher-dimensional analogue.

\subsection{Superpotentials for hyper-K\"ahler geometry}

    In this subsection, we shall show that by imposing an appropriate
set of algebraic restrictions on the four metric functions $a$, $b$, $c$
and $f$, it is possible to reduce the general Lagrangian formulation of
the previous subsection to one that does allow a reformulation in terms
of a superpotential for all dimensions $D=4n+4$, for which the
associated first-order equations imply the special holonomy $Sp(n+1)$ of
hyper-K\"ahler geometry.  Specifically, we impose the two algebraic
equations contained within (\ref{dseq1}) and (\ref{dseq2}), namely
%%%%%
\be
a^2+b^2=c^2\,,\qquad a\, b = c\, f\,.\label{cfsol}
\ee
%%%%%
It is easily verified that these conditions are compatible with the
Einstein equations.  

    Using these conditions to eliminate $c$ and $f$,
we find that for the reduced system where only $a$ and $b$ remain,
the kinetic and potential energies in (\ref{tveq}) become
%%%%%
\bea
T &=& \ft12 g_{ij}\, \fft{\del\a^i}{d\eta}\,
\fft{\del\a^j}{\del\eta}\,,\nn\\
V &=&-\fft{2(a\,b)^{4n}}{(a^2 + b^2)^2}\Big[
(n(2n+1)\, a^8 + 2(4n^2+5n+2)\, a^6\, b^2 \nn\\
&&\qquad+2(6n^2+9n+2)\, a^4\, b^4 + 2(4n^2+5n+2)\, a^2\, b^6
+n(2n+1)\, b^8\Big]\,.
\eea
%%%%%
where $\a^i=(a,b)$, and the new radial variable $\eta$ is now given by
$dt= (a\, b)^{2n}\, c^2\, f\, d\eta = (a\, b)^{2n+1}\, (a^2+b^2)^{1/2}\,
d\eta$.  The sigma-model metric components in $T$ are given by
%%%%%%
\bea
g_{11} &=& \fft{4\Big((n+1)(2n+1)\,a^4 + 2(2n^2+2n+1)\, a^2\,b^2 
+n(2n+1)\,b^4\Big)}{(a^2+b^2)^2}\,,\nn\\
g_{12} &=& \fft{4\Big((n+1)(2n+1)\,a^4 +(4n^2+6n+1)\,a^2\,b^2 
+(n+1)(2n+1)\,b^4\Big)}{(a^2+b^2)^2}\,,\\
g_{22} &=& \fft{4\Big(n(2n+1)\,a^4 +2(2n^2+2n+1)\,a^2\,b^2 
+(n+1)(2n+1)\,b^4\Big)}{(a^2+b^2)^2}\,.\nn
\eea
%%%%%%%%%%
Note that because of the non-linear nature of the algebraic
substitution in (\ref{cfsol}), the sigma-model is now a non-linear one.

   We find that $V$ can now be expressed in terms of a superpotential,
$V=-\ft12g^{ij}\del_i W\,\del_j W$, where $W$ is given by
%%%%%%
\be
W=\fft{2(a\,b)^{2n}}{a^2 +b^2}\, [(2n+1)\, (a^4+b^4) + 4(n+1)\,
a^2\,b^2]\,.
\ee
%%%%%%
The first order equations derived from the superpotential are rather
simple, given (in terms of the original radial variable $t$) by
%%%%%%
\be
\dot a = \fft{b}{\sqrt{a^2 + b^2}}\,,\qquad
\dot b = \fft{a}{\sqrt{a^2 + b^2}}\,.\label{abeq}
\ee
%%%%%%
Despite the rather complicated $n$-dependence of $T$ and $V$, these
first-order equations are independent of $n$.  This should not be
surprising, since we expect that after imposing the algebraic
restrictions (\ref{cfsol}), the solutions should reproduce the Calabi
hyper-K\"ahler metrics, for which the metric functions take the
$n$-independent form (\ref{calsol}).  One can in fact verify that after
substituting (\ref{cfsol}) and (\ref{abeq}) into the general expressions
for the Riemann curvature, the integrability conditions $\Theta_{ab}\,
\Gamma_{ab}\, \eta=0$ imply the existence of $(n+2)$
covariantly-constant spinors, as one expects for hyper-K\"ahler metrics
in $D=4n+4$ dimensions.

   One can of course directly the first-order equations equations
(\ref{abeq}), and one then obtains the hyper-K\"ahler Calabi solutions.
If we introduce $\rho$ by defining $dt=(a^2+b^2)^{1/2}\, d\rho$ then
(\ref{abeq}) are easily solved for $a$ and $b$.  Substituting back to
get the remaining metric functions, we then obtain
%%%%%
\be
a=\sinh\rho\,,\quad b=\cosh\rho\,,\quad
c^2=\cosh 2\rho\,,\qquad f^2= \fft{\sinh^2 2\rho}{4\cosh 2\rho}
\,,\quad h^2=\cosh2\rho\,.
\ee
%%%%%
The further coordinate redefinition $r^2=\cosh2\rho$ gives back the
previous form of the solution,
%%%%%
\be
a^2=\ft12(r^2-1)\,,\quad b^2=\ft12 (r^2+1)\,,\quad c^2=r^2\,,\quad 
f^2 = \ft14 r^2\, (1-r^{-4})\,,\quad h^2 =(1-r^{-4})^{-1}\,.
\label{calabisol}
\ee
%%%%%

\subsection{Superpotentials for Ricci-flat K\"ahler geometry}

   There is a different specialisation of the original $(a,b,c,f)$
system that can be made, which allows the construction of a different
superpotential whose first-order equations imply $SU(2n+2)$ holonomy,
\ie Ricci-flat K\"ahler geometry.  To obtain this case, we need to set
%%%%%
\be
b=a\,,\qquad c=(\sqrt2)\, a\,,\label{bcsol}
\ee
%%%%%
leaving just $a$ and $f$ as dynamical variables.  These conditions are
compatible with the Einstein equations.  In this case the kinetic and
potential energies in (\ref{tveq}) reduce to
%%%%%%%
\bea
T&=& 2(2n+1)(4n+1)\,{\a'}^2 + 4(2n+1)\, \a'\, \sigma'
\,,\nn\\
V&=& -8(2n+1)\, a^{8n}\, f^2\, [2(n+1)a^2 -f^2]\,,
\eea
%%%%%%
where $'$ denotes a derivative with respect to $\eta$, now given by 
$dt=2a^{4n+2}\, f\, d\eta$.  Thus we have
%%%%
\be
g_{ij}=4(2n+1)\pmatrix{4n+1 & 1\cr
                          1 & 0}\,.
\ee
%%%%%%

   We find that this system can be described by a superpotential $W$,
such that $V=-\ft12 g^{ij}\, \del_i W\, \del_j W$, which is given by
%%%%%%
\be
W=4a^{4n}\, [(n+1)\, a^2 + (2n+1)\, f^2]\,.\label{ksupw}
\ee
%%%%%
The associated first-order equations are then 
%%%%%
\bea
\dot a = \fft{f}{a}\,,\qquad
\dot f = \fft{(n+1)\, a^2 - (2n+1)\, f^2}{a^2}\,,\label{afeq}
\eea
%%%%%
when expressed in terms of the original radial variable $t$.  It
should be emphasised that in the special case of $n=1$, where we were
able to obtain the superpotential (\ref{spin7spot}) for Spin(7)
holonomy, the superpotential in (\ref{ksupw}) with $n=1$ is not simply
the one in (\ref{spin7spot}) subject to the constraint (\ref{bcsol}).
The superpotential (\ref{spin7spot}) has an overall factor $f$, but
this is not the case for the superpotential (\ref{ksupw}).

     One can verify from the integrability conditions that these
first-order equations imply the existence of two covariantly-constant
spinors, and hence $SU(2n+2)$ holonomy.  This corresponds to Ricci-flat
K\"ahler geometry.

   It is easy to solve the first-order equations (\ref{afeq}).  If we
define a new radial variable $r$ by $dt=a/(\sqrt2\, f)\, dr$, then the
first equation immediately implies $a=r/\sqrt2$.  It is then
straightforward to solve for $f$.  From these, the rest of the metric
functions follow.  Thus we find
%%%%%
\be
a^2=b^2=\ft12 r^2\,,\qquad c^2= r^2\,,\qquad f^2 = \ft14 U\,
r^2\,,\qquad h^2 = U^{-1}\,,
\ee
%%%%%
where
%%%%% 
\be U= 1- \Big(\fft{r_0}{r}\Big)^{4n+4}\,.
\ee
%%%%%
The resulting Ricci-flat K\"ahler metric in fact lies within the class
constructed in \cite{berber,pagpop}.  It is a
complex line bundle over the homogeneous Einstein-K\"ahler manifold 
$SU(n+2)/(U(n)\times U(1))$.

\section{Harmonic forms in the Calabi metrics}

    In this section, we construct $L^2$-normalisable middle-dimension
harmonic forms in the Calabi metrics of dimension 8 and 12.  Our
principal interest will be in the case of the 8-dimensional
hyper-K\"ahler Calabi metric, since this will allow us to obtain an
explicit solution for a new resolved M2-brane.  By way of an
introduction, we shall begin by presenting the already-known
$L^2$-normalisable harmonic 2-form in the 4-dimensional Calabi metric
(\ie the Eguchi-Hanson solution).  In the subsequent two subsections
following this, we present our new results for the $L^2$-normalisable
harmonic 4-form and 6-form in the hyper-K\"ahler Calabi metrics of
dimensions 8 and 12 respectively.  A fourth subsection contains a
discussion of a regular but non $L^2$-normalisable harmonic 2-form in
the Calabi metrics of arbitrary dimension.

\subsection{$L^2$-normalisable harmonic 2-form in $D=4$}

   The 4-dimensional Calabi metric arises by taking $n=0$ in the
formulae in section 2, in which case (\ref{genmetn}) and
(\ref{calsol}) reduce to the well-known Eguchi-Hanson metric.
In this specific case we adopt the notation that the orthonormal frame
$(e^0,e^{\td1}\, e^{\td2}, e^{\td0})$ of the general formalism in
section 2 will be written simply as $(e^0,e^1,e^2,e^3)$, and so
%%%%%
\be
e^0= h\, dr\,,\qquad e^1= c\, \nu_1\,,\qquad e^2= c\,\nu_2\,,\qquad
     e^3 = f\, \lambda\,.
\ee
%%%%%
The three K\"ahler forms (\ref{3k}) become
%%%%%
\be
J_1=e^0\wedge e^3 + e^1\wedge e^2\,,\qquad 
J_2=e^0\wedge e^1 + e^2\wedge e^3\,,\qquad 
J_3=e^0\wedge e^2 + e^3\wedge e^1\,.\label{d4k}
\ee
%%%%%
There is an $L^2$-normalisable harmonic 2-form, given by
%%%%%
\be
G_\2= \fft1{r^4}\, (e^0\wedge e^3 - e^1\wedge e^2)\,.\label{g2harm}
\ee
%%%%%
Note that in conventions where the K\"ahler forms are self-dual, this
harmonic form is anti-self-dual.  

    We may define an holomorphic complex vielbein $(\ep^0,\ep^1)$ with
respect to any of the three complex structures.  From (\ref{d4k}),
these can be chosen as follows:
%%%%%
\bea
J_1:&& \ep^0 = e^0 + \im\, e^3\,,\qquad \ep^1= e^1 + \im\, e^2\,,\nn\\
J_2:&& \ep^0 = e^0 + \im\, e^1\,,\qquad \ep^1= e^2 + \im\, e^3\,,
\label{d4eps}\\
J_3:&& \ep^0 = e^0 + \im\, e^2\,,\qquad \ep^1= e^3 + \im\, e^1\,.\nn
\eea
%%%%%
In terms of these three different holomorphic bases, we see that the
harmonic 2-form (\ref{g2harm}) is given by
%%%%%
\bea
J_1:&& G_\2= \fft{\im}{2r^4}\, (\ep^0\wedge \bar\ep^0 -
\ep^1\wedge\bar\ep^1)\,,\nn\\
J_2:&& G_\2= \fft{\im}{2r^4}\, (\ep^0\wedge \bar\ep^1 -
\bar\ep^0\wedge\ep^1)\,,\\
J_3:&& G_\2= \fft{1}{2r^4}\, (\ep^0\wedge \bar\ep^1 +
\bar\ep^0\wedge \ep^1)\,,\nn
\eea
%%%%%
Note that $G_\2$ is a type $(1,1)$ form with respect to all three
complex structures $J_i$, and it is orthogonal to all three of the
$J_i$.  This normalisable harmonic 2-form was used to construct resolved
supersymmetric heterotic 5-brane and dyonic string in \cite{clpres}.

\subsection{$L^2$-normalisable harmonic 4-form in $D=8$}

   The structure of the middle-dimension harmonic forms in the
higher-dimensional Calabi metrics is somewhat more complicated.  In
this section, we obtain an explicit expression for an
$L^2$-normalisable harmonic 4-form in the 8-dimensional Calabi metric.

    We begin by considering the 4-form structures that arise from
calculating $\ft12 J_1\wedge J_1$, $\ft12 J_2\wedge J_2$ and $\ft12
J_3\wedge J_3$.  These are in fact all anti-self-dual, in our
conventions.  Then, we shall take these structures, replace the
coefficients ``1'' of each anti-self-dual pair of terms by an arbitrary
function of $r$, and adopt this as our ansatz for an anti-self-dual
harmonic 4-form $G_\4$.  (We may also consider an ansatz for a self-dual
harmonic 4-form where now we first change the previous anti-self-dual
pairs, and then give these arbitrary functions of $r$ as coefficients.
However, this does not give any $L^2$-normalisable harmonic 4-form.)

   From (\ref{d8kahler}), it follows that the ansatz for $G_\4$ will
be
%%%%%
\bea
G_\4 &=& f_1\, (e^0\wedge e^7\wedge e^1\wedge e^2 - 
                  e^3\wedge e^4\wedge e^5\wedge e^6)\nn\\
&& +  f_2\, (e^0\wedge e^7\wedge e^3\wedge e^4 - 
                  e^1\wedge e^2\wedge e^5\wedge e^6)\nn\\
             &&+f_3\, (e^0\wedge e^7\wedge e^5\wedge e^6 - 
                  e^1\wedge e^2\wedge e^3\wedge e^4)\nn\\
&& +  g_1\, (e^0\wedge e^1\wedge e^4\wedge e^5 - 
                  e^2\wedge e^3\wedge e^6\wedge e^7)\label{g4m}\\
             &&+ g_2\, (e^0\wedge e^2\wedge e^3\wedge e^5 - 
                  e^1\wedge e^4\wedge e^6\wedge e^7)\nn\\
&& +  h_1\, (e^0\wedge e^1\wedge e^3\wedge e^6 - 
                  e^2\wedge e^4\wedge e^5\wedge e^7)\nn\\
             &&+ h_2\, (e^0\wedge e^2\wedge e^4\wedge e^6 - 
                  e^1\wedge e^3\wedge e^5\wedge e^7)\,.\nn
\eea
%%%%%
(Note that some structures from $J_2\wedge J_2$ and $J_3\wedge J_3$
just produce a repetition of ones already present from $J_1\wedge
J_1$, and so we have only 7 functions and not 9.)

    The vielbeins are given by (\ref{ortho}), the metric functions are
given by (\ref{calabisol}), and the exterior derivatives of the
various 1-forms in (\ref{ortho}) are given by (\ref{d8ext}).  The
1-form $Q$, which lies outside the $SU(3)/U(1)$ coset, should drop out
of the expressions for $dG_\4$, and this implies the following further
conditions
%%%%%
\be
g_2 = -g_1\,,\qquad h_2=h_1\,.
\ee
%%%%%

   Imposing $dG_\4=0$ now gives a system of five first-order equations
for the remaining five functions.  We find that there is a 
solution regular at $r=1$, given by
%%%%%
\be
f_2 = \fft{1}{r^4\, (r^2+1)}\,,\qquad 
h_1 = \fft{1}{r^2\, (r^2+1)^3}\,,
\ee
%%%%%
together with
%%%%
\be
f_1=f_2+2h_1\,\quad f_3=-2h_1\,,\quad g_1=h_1\,,\quad
g_2=-h_1\,,\quad h_2=h_1\,.\label{linearcon}
\ee
%%%%%
Thus the harmonic 4-form is given by
%%%%%
\crampest
\bea
G_\4 &=& f_2\, (e^0\wedge e^7 -e^5\wedge e^6)\wedge (e^1\wedge e^2 +
e^3\wedge e^4) + 2h_1\, (e^0\wedge e^7 + e^3\wedge e^4) \wedge 
           (e^1\wedge e^2 - e^5\wedge e^6)\nn\\
&&\!\!\!\!\!\!\!\! 
+ h_1\, (e^0\wedge e^6 -e^5\wedge e^7)\wedge (e^1\wedge e^3
+e^2\wedge e^4) \nn\\
&&\!\!\!\!\!\!\!\!
+ h_1\, 
(e^0\wedge e^5 + e^6\wedge e^7)\wedge (e^1\wedge e^4
 -e^2\wedge e^3) \,.\label{g4new}
\eea
\uncramp
%%%%% 
The magnitude of $G_\4$ is given by
%%%%%
\be
|G_\4|^2 = 96( f_2^2 + 2 f_2\, h_1 + 6h_1^2) = 
\fft{96(r^8+6r^6+16r^4+6r^2+1)}{r^8\, (r^2+1)^6}\,.
\label{g4sq}
\ee
%%%%%
It is evident that $G_\4$ is $L^2$-normalisable, since it is regular
at $r=1$ and $|G_\4|^2$ falls off as $1/r^{12}$ at large $r$.
In section 6, we shall use this harmonic form in order to obtain a
new explicit resolved M2-brane solution.

    It is again instructive to re-express the harmonic 4-form in terms
of a complex holomorphic vielbein basis.  As in the previous
subsection, we can use any of the three $J_i$ in order to define the
complex structure.  Let us first consider $J_1$.  From (\ref{3k}),
after changing to our $D=8$ notation, we see that using $J_1$ as the
complex structure we can define the holomorphic vielbeins
%%%%%
\be
\ep^0= e^0+\im\, e^{7}\,,\quad \ep^1= e^{1}+\im\, e^{2}\,,\quad
\ep^2= e^{3}+\im\, e^{3}\,,\quad
\ep^3 = e^{5} + \im\, e^{6}\,.
\ee
%%%%%
It is useful also to define the following real 2-forms:
%%%%%
\be
x_0=\im\, \ep^0\wedge \bar\ep^0\,,\quad
x_1=\im\, \ep^1\wedge \bar\ep^1\,,\quad
x_2=\im\, \ep^2\wedge \bar\ep^2\,,\quad
x_3=\im\, \ep^3\wedge \bar\ep^3\,,
\ee
%%%%%
In terms of these, we find that the harmonic 4-form (\ref{g4new})
is given by 
%%%%%
\bea
G_\4 &=& f\, (x_0-x_3)\, (x_1-x_2) \\
&& + 2g\, [(x_0-x_2)(x_1-x_3)  - \im\, 
\ep^0\wedge \ep^3\wedge \bar\ep^1\wedge\bar\ep^2 + \im\,
\bar\ep^0\wedge \bar\ep^3\wedge \ep^1\wedge \ep^2]\,.\nn
\eea
%%%%%
where the functions $f$ and $g$ are given by
%%%%%
\be
f= \fft1{r^4\, (r^2+1)}\,,\qquad g= \fft1{r^2\, (r^2+1)^3}\,.\label{fg8}
\ee
%%%%%

   If instead we use a complex holomorphic vielbein basis defined by
$J_2$, then we have
%%%%%
\be
\ep^0 = e^0+\im\, e^5\,,\quad \ep^3 = e^6+\im\, e^7\,,\quad 
\ep^1= e^1+\im\, e^4\,,\quad \ep^2 = e^2-\im\, e^3\,.
\ee
%%%%%
In terms of these, we now find that (\ref{g4new}) is given by
$G_\4 = f\, U_1 + g\, U_2$, with $f$ and $g$ given
by (\ref{fg8}) and
%%%%%
\crampest
\bea
U_1 &=& \im\, 
(\ep^0\wedge \bar\ep^3-\bar\ep^0\wedge\ep^3)\wedge 
(\ep^1 \wedge \bar\ep^2 +
\bar\ep^1\wedge \ep^2)\,,\label{g4j2}\\
U_2 &=& U_1 + (x_0+x_3)\, (x_1+x_2) -2 x_1\, x_2 -2 x_0\, x_3
- 2\im\, (\ep^0\wedge \ep^3 \wedge \bar\ep^1\wedge \bar\ep^2
-\bar\ep^0\wedge \bar\ep^3 \wedge \ep^1\wedge \ep^2)\,.\nn
\eea
\uncramp
%%%%%
Finally, we could instead use $J_3$ to define the complex structure.
This gives an expression for $G_\4$ that is very similar to
(\ref{g4j2}).  It is worth remarking that $G_\4$ is of type $(2,2)$ with
respect to all three of the complex structures $J_i$.  It is also
orthogonal to all three $J_i$.

\subsection{$L^2$-normalisable harmonic 6-form in $D=12$}

    The discussion becomes more complicated for the Calabi metric in the
next higher dimension, namely for the harmonic 6-form in $D=12$.  First,
we consider a complex vielbein that is holomorphic with respect to the
complex structure $J_1$:
%%%%%
\bea
&&\ep^0= e^0+\im\, e^{\td0}\,,\quad \ep^1= e^{1,1}+\im\, e^{2,1}\,,\quad
\ep^2= e^{1,2}+\im\, e^{2,2}\,,\nn\\
&&\ep^3= e^{1,\td 1}-\im\, e^{2,\td 1}\,,\quad
\ep^4= e^{1,\td 2}-\im\, e^{2,\td 2}\,,\quad
\ep^5 = e^{\td1} + \im\, e^{\td 2}\,.
\eea
%%%%%
We also define the following (real) 2-forms:
%%%%%
\be
x_0=\im\, \ep^0\wedge \bar\ep^0\,,\quad
x_1=\im\, \ep^1\wedge \bar\ep^1\,,\quad
x_2=\im\, \ep^2\wedge \bar\ep^2\,,\quad
x_3=\im\, \ep^3\wedge \bar\ep^3\,,\quad
x_4=\im\, \ep^4\wedge \bar\ep^4\,,\quad
x_5=\im\, \ep^5\wedge \bar\ep^5\,.
\ee
%%%%%

   We then proceeded by making a rather general ansatz for the 6-form,
with as-yet undetermined functions of $r$ as coefficients for each
structure in the ansatz.  After calculations of some complexity, we
find after imposing (anti) self-duality, and $dG_\6=0$, that there is
an $L^2$-normalisable harmonic 6-form, given by
%%%%%
\bea
G_\6 &=& f\, (x_0-x_5)\, [3 x_1\, x_2 + 3 x_3\, x_4 -x_1\, x_3 - 
x_2\, x_4 - 2x_1\, x_4 - 2x_2\, x_3\nn\\
&& - \ep^1\wedge \bar\ep^2\wedge \ep^3\wedge \bar \ep^4 -
\bar\ep^1\wedge \ep^2\wedge \bar\ep^3\wedge \ep^4)]\nn\\
&&+ 2g\, [S -(x_0-x_5)\, ( \ep^1\wedge \bar\ep^2\wedge \ep^3\wedge \bar
\ep^4  +\bar\ep^1\wedge \ep^2\wedge \bar\ep^3\wedge \ep^4) \nn\\
&&-3\im\, (x_1-x_3)\, (\ep^0\wedge\ep^5\wedge \bar\ep^2\wedge\bar\ep^4 
                     -\bar\ep^0\wedge\bar\ep^5\wedge\ep^2\wedge\ep^4)\nn\\
&&-3\im\, (x_2-x_4)\, (\ep^0\wedge\ep^5\wedge \bar\ep^1\wedge\bar\ep^3 
                     -\bar\ep^0\wedge\bar\ep^5\wedge\ep^1\wedge\ep^3)]\,,
\label{6form}
\eea
%%%%%
where
%%%%%
\bea
S &\equiv& (x_0-x_5)\, (3x_1\, x_2 + 3 x_3\, x_4 - x_1\, x_3 - x_2\,
x_4 - 2x_1\, x_4 - 2x_2\, x_3) \nn\\
&& + 3 (x_0+x_5) (x_1\, x_2 -x_3\, x_4) 
- 3 x_0\, x_5\, (x_1+x_2-x_3-x_4)\nn\\
&& -3x_1\, x_2\,( x_3 +x_4) 
      + 3 x_3\, x_4\, (x_1+x_2)\,.\label{Qdef}
\eea
%%%%%
The two functions $f$ and $g$ are given by
%%%%%
\be
f= \fft1{r^4\, (r^2+1)^2}\,,\qquad g= \fft1{r^2\, (r^2+1)^4}\,.
\label{fg12}
\ee
%%%%%
The harmonic 6-form is normalisable, and anti-self-dual.  It is
orthogonal to all three K\"ahler forms $J_i$, and it is manifest from
(\ref{6form}) and (\ref{Qdef}) that it is of type $(3,3)$.  The
magnitude of $G_\6$ is given by
%%%%
\be
|G_\6|^2=15 \times 6!\, (f^2 + 4f\, g + 16 g^2)\,.
\ee
%%%%%

   We may instead express the harmonic 6-form in terms of a complex
basis that is holomorphic with respect to the complex structure $J_2$.
From (\ref{3k}), we may choose
%%%%%
\bea
&&\ep^0=e^0+\im\, e^{\td1}\,,\quad \ep^5=e^{\td0}-\im\, e^{\td2}
\,,\quad \,,\nn\\
&&\ep^1= e^{1,1} + \im\, e^{2,\td1}\,,\quad
 \ep^2= e^{1,2} + \im\, e^{2,\td2}\,,\quad
\ep^3= e^{2,1} - \im\, e^{1,\td1}\,,\quad
 \ep^4= e^{2,2} - \im\, e^{1,\td2}\,.
\eea
%%%%%
In terms of this basis, we find $G_\6= f\, U_1 + g\, U_2$ with $f$ and
$g$ given again by (\ref{fg12}), and
%%%%%
\bea
U_1 &=& \ep^0\wedge \bar\ep^5 \wedge [(x_1+x_3)(x_2+x_4) 
 -2x_1\, x_3 -2 x_2\, x_4 -5P\nn\\
&&\qquad\qquad -2\ep^1\wedge \bar\ep^2\wedge \ep^3\wedge\bar\ep^4 
 - 2\bar\ep^1\wedge \ep^2\wedge \bar\ep^3\wedge\ep^4]
+\hbox{c.c.}\,,\nn\\
U_2 &=& U_1 - 6\ep^0\wedge\ep^5\wedge S -6\bar\ep^0\wedge\bar\ep^5\wedge
\bar S\nn\\
&&+ 3(\ep^1\wedge\bar\ep^3+\bar\ep^1\wedge\ep^3)\wedge 
  [(x_0+x_5)(x_2+x_4) -2x_2\, x_4 -2 x_0\, x_5]\nn\\
&& +3(\ep^2\wedge\bar\ep^4+\bar\ep^2\wedge\ep^4)\wedge 
  [(x_0+x_5)(x_1+x_3) -2x_1\, x_3 -2 x_0\, x_5]\,,
\eea
%%%%%
where c.c. indicates that the complex conjugate of the entire
expression presented for $U_1$ is to be added, and here we have
defined 
%%%%%
\bea
P &\equiv& \ep^1\wedge \ep^2\wedge \bar\ep^3\wedge\bar \ep^4 +
\bar\ep^1\wedge \bar\ep^2\wedge \ep^3\wedge\ep^4
+ \ep^1\wedge \bar\ep^2\wedge \bar\ep^3\wedge\ep^4
+ \bar\ep^1\wedge \ep^2\wedge \ep^3\wedge\bar \ep^4\,,\nn\\
S &\equiv& \ep^1\wedge \bar\ep^2\wedge \bar\ep^3\wedge\bar \ep^4 +
\bar\ep^1\wedge \ep^2\wedge \bar\ep^3\wedge\bar\ep^4
+ \bar\ep^1\wedge \bar\ep^2\wedge \ep^3\wedge \bar\ep^4
+ \bar\ep^1\wedge \bar\ep^2\wedge \bar\ep^3\wedge \ep^4\,.\nn\\
\eea
%%%%%
Note that $G_\6$ is of type $(3,3)$ in this basis also.

\subsection{A non-normalisable harmonic 2-form}

    Consider the $U(1)$ Killing vector dual to the direction of the
1-form $\lambda$.  If we lower the index, we get a Killing 1-form
%%%%%
\be
A=f^2\, \lambda\,.
\ee
%%%%%
Because it is a Killing 1-form, and the metric is Ricci-flat, we have that
$d{*dA}=0$, as well, of course, as $ddA=0$, and so $dA$ is harmonic.
We find that
%%%%%
\bea
dA &=& \fft{2 f'}{h}\, e^0\wedge e^{\td0} + \fft{2f^2}{a^2}\,
e^{1\a}\wedge e^{2\a} - \fft{2f^2}{b^2}\, 
e^{1\td\a}\wedge e^{2\td\a} + \fft{4f^2}{c^2}\,
e^{\td1}\wedge e^{\td2}\,,\nn\\
&\equiv & J_1 + G_\2\,,
\eea
%%%%%
where $J_1$ is the first K\"ahler form in (\ref{3k}), and
%%%%%
\be
G_\2 = \fft1{r^4}\, (e^0\wedge e^{\td0} - e^{\td1}\wedge e^{\td2}) +
\fft1{r^2}\, (e^{1\a}\wedge e^{2\a} + e^{1\td\a}\wedge e^{2\td\a})\,.
\label{g2gen}
\ee
%%%%%
Since we know that $J_1$ itself, being a K\"ahler form, is harmonic, and
we know that $dA$ is harmonic, it follows that $G_\2$ is harmonic.  It
is not in general $L^2$ normalisable, since $|G_\2|^2$ is proportional
to $r^{-4}$, but it is regular and square-integrable at $r=1$.  Note
that in the special case of $D=4n+4=4$ dimensions, for which the Calabi
metric is just the Eguchi-Hanson solution, the terms proportional to
$1/r^2$ in (\ref{g2gen}) are absent and so $G_\2$ reduces to precisely
the $L^2$-normalisable harmonic 2-form given in (\ref{g2harm}) in this
case.

\section{A resolved M2-brane in the Calabi 8-metric}

\subsection{The M2-brane solution, and its supersymmetry}

    In section 5.2 we constructed an $L^2$-normalisable anti-self-dual
harmonic 4-form (\ref{g4new}) in the eight-dimensional hyper-K\"ahler
Calabi metric.  We may use this in order to obtain a new resolved
M2-brane solution.

    As a consequence of the Chern-Simons modification to the equation of
the motion of the 3-form potential in $D=11$ supergravity, namely
%%%%%%
\be
d\hat {* F_\4} = \ft12 F_\4\wedge F_\4\,.
\ee
%%%%%
it is possible to construct a deformed M2-brane, given by \cite{hawtay,clpres}
%%%%%%%%
\bea 
d\hat s_{11}^2 &=& H^{-2/3}\, dx^\mu\, dx^\nu\, \eta_{\mu\nu} +
H^{1/3}\, ds_8^2\,,\nn\\
 F_\4 &=& d^3x\wedge dH^{-1} + m\, G_\4\,, \label{d11sol}
\eea 
%%%%%
where $G_\4$ is the harmonic anti-self-dual 4-form in the Ricci-flat
transverse space $ds_8^2$, and the function $H$ satisfies
%%%%%%%
\bea
\square H = -\ft1{48} m^2\, G_\4^2\,.\label{d2sol}
\eea
%%%%%
Warped reductions of this type, were also discussed in 
\cite{gss,2beckers,becker}. 

    Using the expression (\ref{d8calmet}) for $ds_8^2$, and
(\ref{g4sq}) for $|G_\4^-|^2$, it follows that the solution for the
metric function $H$ in the corresponding resolved M2-brane is
%%%%%
\bea
H &=& c_0 + \fft{m^2}{48}\, \int_r^\infty dr'\, \Big(\fft{h^2}{\sqrt{g}}
\, \int_1^{r'} dr''\, \sqrt{g}\, |G_\4|^2\Big)\nn\\
&=& c_0 + \fft{m^2\, (5r^6 + 25 r^4 + 48 r^2 + 40)}{160 r^2\,
(r^2+1)^5}\,.\label{Hfun}
\eea
%%%%%   
At small distance, \ie  $r\rightarrow 1$, the function $H$ tends to a
constant.  At large $r$, $H$ has the asymptotic behaviour
%%%%%
\be
H\sim c_0 + \fft{m^2}{32 r^6} - \fft{m^2}{80 r^{10}} - \fft{m^2}{32
r^{14}} +\cdots\,.
\ee
%%%%%
In terms of proper distance $\rho$ defined by $d\rho=h\, dr$, we have
%%%%%
\be
H\sim c_0 + \fft{m^2}{32 \rho^6} - \fft{7m^2}{160 \rho^{10}} +
\fft{3m^2}{224 \rho^{14}} +\cdots\,.
\ee
%%%%%%%
In the decoupling limit, where $c_0$ can be dropped, the resolved
M2-brane approaches AdS$_4\times \Sigma_7$ at large distance, where
$\Sigma_7$ is the $SU(3)/U(1)$ Einstein space known as the $N^{1,1}$
manifold.  The solution therefore interpolates between this geometry at
large distance and a geometry that is locally $M_3\times R^4\times S^4$
at short distance, where $M_3$ denotes 3-dimensional Minkowski spacetime.

    We now turn to a discussion of the supersymmetry of our new
resolved M2-brane solution.  The criterion for supersymmetry is
%%%%%
\be
G_{abcd}\, \Gamma_{bcd}\, \eta=0\,,\label{susycon}
\ee
%%%%%
where $\eta$ is a covariantly-constant spinor in the 8-dimensional
Ricci-flat metric.  Substituting (\ref{g4new}) into this, and making use
of the conditions (\ref{gammacon}) that define the three
covariantly-constant spinors in the 8-dimensional Calabi metric, we find
that (\ref{susycon}) is satisfied by all three of the
covariantly-constant spinors.  To be precise, the constraint
(\ref{susycon}) is consistent with the linear relations amongst the
coefficients of the harmonic 4-form given in (\ref{linearcon}).  Thus
our solution is a supergravity dual of an ${\cal N}=3$ conformal field
theory.

        It is possible to perform a dimensional reduction of the 
M2-brane on the principal orbits $N^{1,1}$ of the transverse space, 
and thereby obtain a four-dimensional domain wall, given by
%%%%%%%
\be
ds_4^2 = r^7(1-r^{-4})^{3/2}\, \Big[
H^{1/2}(-dt^2 + dx_1^2 + dx_2^2) + H^{3/2}\, \fft{dr^2}{1-r^{-4}}
\Big]\,,
\ee
%%%%%%
which is asymptotically AdS$_4$.  It is easy to verify that the minimally
coupled scalar wave equation in this background has a discrete spectrum,
indicating the confinement of the dual conformal gauge theory.  We shall
discuss the properties of the dual field theories in more detail in the
following section.

\subsection{General features of resolved brane solutions}

   It is worth pausing at this point to make some observations about
certain general features of the resolved M2-brane solution, and their
relation to corresponding properties of harmonic 4-form.  In fact
although we shall make these remarks in the specific context of the
resolved M2-brane, they are quite general, and apply equally well to
any resolved brane solution in which an additional field strength is 
taken to be proportional to an harmonic form.  

    There are two characteristics of the harmonic form that are
relevant.  Firstly, there is the question of regularity, \ie whether
or not the form is well-behaved at the ``origin,'' which lies at $r=1$
in our present example.  Secondly, there is the issue of
normalisability, and in particular, whether the harmonic form is
square-integrable at large distance.  If the form is in $L^2$, such as
the 4-form that we have obtained in this paper, then it satisfies both
the conditions of short-distance regularity and large-distance
square-integrability.  In other examples, such as the harmonic 3-form
in the deformed fractional D3-brane solution of \cite{klebstra}, it is
regular at the origin but it is not square-integrable at large
distance.  In yet further examples, such as the fractional D3-brane
using the ``small resolution'' of \cite{tz1,tz2}, there is a harmonic form
that is neither regular at short distance nor square-integrable at
large distance.

    The regularity of $G_\4$ at short distance implies that the lower
limit in the $r''$ integral in (\ref{Hfun}) can be taken to be the
``origin'' at $r=r_0=1$, which ensures that the solution (\ref{d11sol})
has no naked singularity at $r=1$.  The $L^2$-normalisability of the
$G_\4$ ensures that the electric M2-brane charge is then a
(non-vanishing) finite constant, given by
%%%%%%
\be
Q=m^2\, \int_{r_0}^\infty dr\, \sqrt{g}\, |G_\4|^2\,.
\ee
%%%%%
By contrast, in the deformed fractional D3-brane of \cite{klebstra}
the analogous D3-brane ``charge'' grows logarithmically with radius,
as a consequence of the lack of large-distance square-integrability.
The solution is, however, still regular at short distance.  On the
other hand in the small resolution of the D3-brane \cite{tz1,tz2}, as well
as the logarithmic growth of the ``charge'' resulting from the lack of
square-integrability at large distance, there is also a ``repulson''
singularity at short distance, resulting from the short-distance
irregularity of the harmonic 3-form in that example.

\section{Comments on dual gauge three-dimensional theories}

\subsection{M2-brane/CFT$_3$ correspondence}

      Having obtained the resolved M2-brane on the 8-dimensional
hyper-K\"ahler Calabi metric, it is of interest to study this
solution, and others, from the point of view of the AdS$_4$/CFT$_3$
correspondence.  A large class of explicit M2-brane solutions on
various supersymmetric 8-dimensional manifolds was obtained in
\cite{clpres,cglp1,cglp2}.  The properties of the 8-dimensional
manifolds are tabulated in appendix A.  In this section we shall make
comments on properties of the associated dual field theories as well.

\subsubsection{Symmetry groups of the field theory}

       The global symmetries of the matter sector of the dual gauge
theory is determined by the symmetry of the Ricci flat manifold,
specifying the coordinates of the transverse space of the $p$-brane.
Namely, the matter multiplets form representations of the global
symmetry group of the SYM theory, which is isomorphic to the isometry
group of the transverse space. These multiplets are expected to transform
as bi-fundamentals of the SYM gauge factors, in accordance with the
quiver theory that specifies the world-volume theory of the resolved
brane. The global symmetry groups and the other properties of the
various M2-branes constructed in this and the previous papers
\cite{clpres,cglp1,cglp2} are summarised in the following table.

\bigskip\bigskip
\centerline{
\begin{tabular}{|c|c|c|c|c|}\hline
Susy      &    $K$      & Global   &    Flow Para.\\
          &             & Symmetry & $\gamma$ 
 \\   \hline\hline
${\cal N}=1$ & $S^7$  &$SO(5)\times SO(3)$ & $\ft43$\\ \hline\hline
%%%
${\cal N}=0$     & $S^7$  & $U(4)$ &   8\\ \hline
${\cal N}=0$  
 & $N^{1,1}$ & $U(3)$ & 8 \\ \hline
${\cal N}=2$ & $M^{2,3}$ & $SU(3)\times U(2)$ &4 \\ \hline
 ${\cal N}=2$  
&$Q^{1,1,1}$ & $SU(2)^3\times U(1)$ &  4\\ \hline
 ${\cal N}=2$  
&$Q^{1,1,1}$ & $SU(2)^3\times U(1)$ & 
 4\\ \hline\hline
%%%%%
${\cal N}=3$ & $N^{1,1}$ & $U(3)$ &
 4\\ \hline
\end{tabular}}
\bigskip

\noindent{Table 1:} Global symmetries of the transverse spaces,
determining the global symmetries of the dual three-dimensional gauge
theories for resolved M2-branes.  The ${\cal N}=1$ solution uses the
8-manifold of Spin(7) holonomy.  The ${\cal N}=0$ and ${\cal N}=2$
examples use Ricci-flat K\"ahler 8-manifolds.  The example with
$K=Q^{1,1,1}$ is listed twice because there are two different
resolutions, one with $S^2\times S^2$ degenerate orbits, and the other
with $S^2\times S^2\times S^2$.  In the ${\cal N}=3$ example the
transverse space is the 8-dimensional hyper-K\"ahler Calabi metric.

\bigskip\bigskip

   It is worth remarking that the two ${\cal N}=0$ cases in Table 1
arise because the associated harmonic 4-forms do not satisfy the
supersymmetry criterion (\ref{susycon}), even though the 8-metrics
themselves admit two covariantly-constant spinors.  Thus in these two
cases the procedure of breaking the conformal symmetry by resolving
the cone metric simultaneously leads to a breaking of supersymmetry.

    The ``flow parameter'' $\gamma$ in Table 1 measures the deviation
from an asymptotically AdS$_4$ behaviour at large proper distance
$\rho$.  It is defined by looking at the behaviour of the metric
function $H$ in (\ref{d11sol}) at large distance, in terms of the
proper-distance coordinate $\rho$:
%%%%%
\be
H = c_0+ \fft{Q}{\rho^6}(1 -\fft{c}{\rho^{\gamma}} + \cdots)\,.
\ee
%%%%%

      The actual determination of the local gauge symmetry and the
spectrum of the matter super-multiplets, in the absence of reliable
field theory calculations (such as in the case of $S^5/\Z_2$ orbifold
blow-up, which leads, in the infra-red, to the SYM theory on the conifold
(see \cite{klewit} in four-dimensional field theory)), often relies on
symmetry and an intuitive approach.  In certain cases, this is enough.  For
example, the isometry group of the transverse space for AdS$_5\times S^5$
is $SO(6)\times U(1)$, and the six scalars form an irreducible (vector)
representation of $SO(6)$.  It follows that all the six scalars must
carry the same representation of the local gauge group, namely $SU(N)$.
For more complicated cases such as our resolved branes, such a
determination has so far only been achieved for D3-branes or M2-branes
in which the transverse-space manifolds are of the toric variety.  In these
cases the geometric constraints of the conifold allow one to determine
the so-called D-term constraints and fix the superpotential interactions
for the massless matter supermultiplets.  For the $T^{1,1}$ conifold,
the local gauge group turns out to be $SU(N)\times SU(N)$, which can be
enhanced to $SU(N)\times SU(N+M)$ when additional fractional D5-branes
wrap around the 2-cycle of the $T^{1,1}$ space \cite{gubkle}.
The D-terms for the $Q^{1,1,1}$ and $M^{2,3}$ spaces were determined in
\cite{ohtat} and \cite{dall} respectively.  From these D-term
constraints, the gauge groups can be shown to be $SU(N)\times
SU(N)\times SU(N)$ and $SU(N)\times SU(N)$ respectively \cite{ffgrtzz}.

\subsubsection{Brane resolution and confinement}

       In the resolved M2-brane construction there are two stages to
the resolution. The first stage consists of resolving the cone manifold
(\ie conifold) itself, and replacing it with a complete Ricci-flat
manifold that asymptotically approaches the original conifold at large
distance.  Having done this, the usual construction of an M2-brane is
typically singular at small distance.  At the second stage the
singularity is resolved by setting the $F_\4$ field strength equal to a
regular harmonic self-dual 4-form in the transverse 8-metric.  If the
harmonic 4-form is normalisable as well, then the AdS structure is
preserved at large distance. In this case, in the ultraviolet regime 
(the gravity dual at large distance) the dual field theory approaches
a three-dimensional conformal field theory with less than maximal 
supersymmetry.

    In the case of the M2-brane, with an 8-dimensional transverse
space that has the asymptotic structure of a cone metric, there are no
known examples where in addition M5-branes can wrap around
supersymmetric 3-cycles \cite{herkle}.  (First-order equations for
wrapped M5-branes on various supersymmetric cycles, without the
presence of the M2-brane, were obtained in \cite{gkw}, following the
construction of an NS-NS 5-brane wrapped on a 2-cycle \cite{malnun}.)
Thus all the resolved M2-branes constructed in this paper and in
\cite{clpres,cglp1,cglp2} carry only electric charge for the $F_\4$
field strength, and the magnitude of this charge is governed by the
magnitude of the harmonic 4-form.  As a consequence, the dual field
theory in the infrared (the gravity dual at small distance) is
described as a perturbation of the CFT by relevant operators,
identified with pseudo-scalar fields of the dual field theory
\cite{herkle}.

     One might wonder why it is necessary to consider the resolution of
the solution, since although the original cone metric is singular at the
origin, this singularity is smoothed by the presence of the M2-brane
located on it, which implies that the geometry near the cone singularity
becomes AdS$_4\times \Sigma_7$, where $\Sigma_7$ is a smooth Einstein
manifold.  This solution might seem already to be a perfectly satisfactory
supergravity dual for the three-dimensional conformal field theory with
reduced supersymmetry, since the solution is regular everywhere.
However, it is easy to verify that the minimally-coupled scalar wave
equation for the AdS$_4$ horospherical metric,
%%%%%%
\be
ds_{{\rm AdS}_4}^2 = \fft1{z^2}\, (-dt^2 +dx_1^2 + dx_2^2 + dz^2)\,,
\ee
%%%%
has a continuous spectrum with no mass gap.  To be specific, although
the UV boundary provides an infinite wall to the Schr\"odinger
potential, it vanishes in the IR region where $z\rightarrow\infty$.  
Thus there would be no indication of confinement in the dual gauge theory.

       The picture is rather different for the resolved M2-branes.
First of all, the geometry of the resolved conifold itself becomes
$\R^n\times \Sigma_{8-n}$ at small distance, where $\rho$ is the proper
distance from the origin.  The usual construction of the M2-brane would
imply that we have $H=1/\rho^{n-2}$ at small distance, which would then
result in a singularity.  However, if the 8-manifold has a harmonic
self-dual 4-form $G_\4$ that is regular at short distance, and we set
$F_\4=m\, G_\4$, then with the integration constants chosen as in
(\ref{Hfun}), $H$ will approach a non-vanishing constant at small
$\rho$, and hence the solution becomes regular there.  The
geometry in the infrared region then becomes a smooth manifold
%%%%%%
\be M_3\times R^n\times \Sigma_{8-n}\,,  
\ee 
%%%%%
where $M_3$ denotes 3-dimensional Minkowski spacetime.
This implies that the transverse space at short distance tends to the
finite values of the metric coefficients, rather than an AdS horizon.
In turn, this means that the Schr\"odinger potential for the
minimally-coupled scalar wave equation is of the infinite-well type, and
hence that the spectrum is discrete, indicating confinement of the
corresponding gauge theory.

    The decoupling limit of the resolved M2-brane interpolates between
$M_3\times R^n\times \Sigma_{8-n}$ in the IR region and AdS$_4\times
\Sigma_7$ at the UV boundary region.  Owing to the fact that the
harmonic 4-form $G_\4$ does not carry a non-vanishing flux for the
M5-brane charge, one expects that the flow from the conformal point of
AdS$_4\times \Sigma_7$ to a non-conformal $M_3\times R^n\times
\Sigma_{8-n}$ will be described by a perturbation by relevant operators,
associated with the pseudoscalar fields of the gauge theory in the
Higgs branch.  Thus we see that these relevant operators play an
important r\^ole in the context of confinement.  This is very
different from the D3-brane resolution, and hence the mechanism for
confinement for the ${\cal N}=2$, $D=4$ gauge theory, where the additional
fractional wrapped D5-brane flux is needed \cite{klebtsey,klebstra}.

\subsection{D2-brane/QFT$_3$ correspondence}

         The D2-brane differs from the M2-brane in two important
respects.  First of all, the D2-brane does not have an asymptotic AdS
structure, and hence the dual gauge theory is not a conformal one.
Secondly, it is possible to wrap additional branes, which can be either
D4-branes or NS-NS 5-branes, on non-trivial cycles in the 7-dimensional
transverse space.  The properties of some non-trivial complete
Ricci-flat 7-manifold are summarised in appendix A.  The D2-branes
with wrapped fractional D4-branes or NS-NS 5-branes in these manifolds
were constructed in \cite{clpres,cglp2}, and they were shown to preserve
${\cal N}=1$ supersymmetry \cite{cglp2}.  Here, we list the global
symmetries of the transverse spaces, which in turn determine the global
symmetries of the spectrum in the dual field theory. 

\bigskip
\centerline{
\begin{tabular}{|c|c|c|}\hline
$K$ & Brane & Global  %& $g_1^{-1}-g_2^{-1}$
\\
    & Charge&Symmetry %&
\\ \hline\hline
$\CP^3$ & $D2(N), D4(M)$ & $SO(5)\times SO(3)$
  \\ \hline
Flag$_6$& $D2(N), D4(M)$ & $SU(3)\times SU(2)$
%& $\sim \fft{1}{r}$
 \\
  \hline
$S^3\times S^3$ &$D2(N), NS5(M)$ & $SO(4)\times SO(3)$
%& $\sim r$
\\ \hline
\end{tabular}}

\bigskip
\noindent{Table 2:}  Global symmetries of
 dual  three dimensional field theories of
resolved D2-branes.
\bigskip\bigskip 

    We should like at this point 
to venture into trying to address the spectrum of
the dual field theory, in particular for the resolved D2-brane with the
base $S^3\times S^3$, for which the topology of the transverse space is
$\R^4\times S^3$.  The topology can be
encoded by embedding this space into $R^8$, and imposing the constraint
\cite{amv}:
%%%%%%
\be
\sum_{i=1}^4 x_i^2 -\sum_{j=1}^4 y_i^2 =r>0\,, \label{const}
\ee
%%%%%%%%
where  the $x_i$ coordinates parameterise the $S^3$. Reinterpreting this
constraint in terms of quaternionic variables \cite{amv}:
$\{ x_i\}\rightarrow   q_1=\sum_i u_i\sigma_i$ for  $\{y_i\}\rightarrow
q_2= \sum_i v_i\sigma_i$, where $\sigma_i$ are the quaternionic generators,
allows one to rewrite it as:
%%%%%%
\be
|q_1|^2-|q_2|^2=r\,.
\ee
%%%%% 
The global symmetry $[SU(2)_1\times SU(2)_2]_L \times [SU(2)_{1+2}]_R$
of the manifold acts on the quaternionic fields $q_{1,2}$ by the left
and right $[SU(2)_{1,2}]_{L,R}$ multiplications, respectively.  One
may attempt to identify these fields with the matter fields of the
dual theory, and they transform under the above global symmetry as:
$q_1=({\bf 2},{\bf 1},{\bf 2})_{N,{\bar N}+{\bar M}}$, $q_2=({\bf 1},
{\bf 2},{\bf 2})_{{\bar N}, N+ M}$.  Of course the conjectured form of
the matter spectrum would have to be confirmed by independent
calculation. Here we have also tacitly assumed that the gauge group
of the theory is $SU(N)\times SU(N+M)$.

   The dual field theory has ${\cal N}=1$ supersymmetry, and as a
consequence the assumed interaction terms for the proposed matter
fields would be of the form
%%%%%
\be
W=\epsilon_{BD}\epsilon_{B'D'}\epsilon_{AA'}\epsilon_{CC'}
[q_{1}]_{AB} [q_{2}]_{CD} [q_{1}]_{A'B'}
[q_{2}]_{C'D'} \,,\label{interaction}
\ee
%%%%%%%
where an implicit summation over gauge indices is understood.

 This fractional D2-brane is obtained if the transverse
seven-dimensional manifold has a non-trivial 3-cycle, around whose
dual 3-cycle an NS-NS 5-brane can wrap.  In this case, we take
$G_\3=\omega_\3$, where $\omega_\3$ is the harmonic form 
associated with the 3-cycle.  The function $H$ in the M2-brane 
metric (\ref{d11sol}) is given at large distance by \cite{cglp2}
%%%%%
\be
H= c_0 + \fft{m^2}{4 r^4} + \fft{Q}{r^5}+\cdots \,.
\ee
%%%%%
There is now a term with the slower fall-off $1/r^4$ than the usual
$1/r^5$ term for the standard M2-brane, owing to the fact that
$\omega_3$ is not $L^2$ normalisable at large $r$.  In particular, the
deformed solution no longer has a well-defined ADM mass.  In other
words, the effective 4-form ``electric charge'' is proportional to
$r^6\, H'$ which is $\propto m^2\, r$ thus confirming the
indication in the dual field theory that the difference of the
inverses of the two gauge-group factors, $g_1^{-2}-g_2^{-2}$, is
proportional to $M\sim m^2$.  As should be the case in
three-dimensional field theories, this difference grows linearly with
energy (which on the gravity side it proportional to the distance
$c$).

    On the other hand fractional D2-brane solutions associated with
the $\CP^3$ or ${\rm Flag}_6$ base manifolds arise from the existence
of a non-trivial 4-cycle around whose dual 2-cycle a D4-brane can wrap
\cite{cglp2}.  If $\omega_\4$ denotes the associated harmonic 4-form
in the transverse 7-space, we can set $G_\4=\omega_\4$ in
(\ref{d2sol}) in order to obtain a solution.  One finds that the
metric function $H$ in the M2-brane solution (\ref{d11sol}) 
is given at large distance, for a suitable
normalisation for $\omega_\4$, by \cite{herkle}
%%%%%
\be
H = c_0 + \fft{Q}{r^5} - \fft{m^2}{4 r^6}+\cdots \,.\label{con1h}
\ee
%%%%%
The fractional D2-brane carries an electric charge $Q$ and a magnetic
charge $m$ for $F_\4$, while the 3-form, given by $F_\3= m\, r^{-4}\,
dr\wedge {*_{\sst 6}\omega_\4}$, has vanishing flux integral.  This
result indicates that in contrast to the previous example, here in the
dual three-dimensional theory the leading contribution to the running
of the difference of gauge couplings is absent.

       When the transverse space is flat, corresponding to the
maximally-supersymmetry case, the D2-brane can be viewed as a periodic
array of isotropic M2-branes, and hence the usual D2-brane is a limit of
the M2-brane \cite{imjy}.  Such a connection between the M2-brane and
the D2-brane becomes much less obvious in the non-trivial 7-geometries.
The supersymmetry suggests that the fractional D2-brane might be related
to an M2-brane resolved on an 8-manifold of Spin(7) holonomy.  We leave
this for future investigation. 

\section{Conclusions}

    In this paper, we have given an explicit construction of the
hyper-K\"ahler Calabi metrics, for all dimensions $D=4n+4$.  This uses
the fact that the metrics have cohomogeneity one, with principal orbits
described by the coset $SU(n+2)/U(n)$.  We began by making an ansatz for
such metrics, with undetermined functions of the radial coordinate as
scaling factors for the various terms in the homogeneous level sets.
Following the procedure of Dancer and Swann \cite{danswa}, we then
obtained first-order equations for these functions as integrability
conditions for the existence of a hyper-K\"ahler structure.  Our
solution of these equations gives very simple expressions for
the hyper-K\"ahler Calabi metrics.

   Having obtained the Calabi solutions, it is of interest also to see
how they can arise as the solutions of a system of first-order
equations derivable from a superpotential.  In fact from a Lagrangian
formulation of the Einstein equations for our original metric ansatz
we were able in general to derive two inequivalent superpotential
descriptions; one leading to the hyper-K\"ahler solutions, and the
other to Ricci-flat K\"ahler solutions on the complex line bundles
over $SU(n+2)/(U(n)\times U(1))$.

   The Calabi metric in dimension $D=4n+4$ is defined on a complete
non-compact manifold that is asymptotic to the cone over $SU(n+2)/U(n)$.
Thus it can be viewed as a resolution of the conifold where the base of
the cone is $SU(n+2)/U(n)$.  In an appendix, we studied this conifold,
using a quaternionic construction in which one starts from $\H^{\, n+2}$
with coordinates $q_A$ and imposes the quadratic constraint $q_A^T\,
q_A=0$, which imposes three real conditions, together with
$q_A^\dagger\, q_A=1$ to set the radius of the cone to unity.  We showed
how the Einstein metric on the $SU(n+2)/U(n)$ base of the cone arises as
the restriction of the flat metric on $\H^{\,n+2}$ under the quadratic
constraints, together with a further $U(1)$ Hopf fibration.  This is
the hyper-K\"ahler quotient construction, which we then applied also
in the non-singular case, thereby obtaining an alternative derivation
of the Calabi metrics in the form (\ref{calabimetric}). We also
showed how this contrasts with the complex conifold construction, where
the base of the cone is instead $SO(n+2)/SO(n)$.  In this case the
analogous constraints on the coordinates of $\C^{n+2}$ directly yield
the manifold of the $SO(n+2)/SO(n)$ base, with no need for a further
Hopf reduction.  However, in this example the Einstein metric on
$SO(n+2)/SO(n)$ is not simply the metric inherited from the flat metric
on $\C^{n+2}$; instead, it is obtained by adding an additional invariant
term which corresponds to a ``squashing'' of the Hopf fibres in
$SO(n+2)/SO(n)$ viewed as a $U(1)$ bundle over the real Grassmannian
$SO(n+2)/(SO(n)\times SO(2))$.

   One of our motivations for obtaining the explicit construction of
the hyper-K\"ahler Calabi metrics was in order to allow us to solve
explicitly for $L^2$-normalisable middle-dimension harmonic forms.  In
particular, our principle interest was in the 8-dimensional Calabi
metric, which can be used in order to obtain a resolved M2-brane
solution of $D=11$ supergravity.  We therefore solved explicitly for
the $L^2$-normalisable harmonic 4-form in the 8-dimensional Calabi
metric.  We also obtained the result for the normalisable harmonic
6-form in the 12-dimensional Calabi metric.

   Using the harmonic 4-form in the 8-dimensional Calabi metric, we
then constructed a deformed M2-brane solution.  We showed that all
three of the covariantly-constant spinors in the Calabi metric remain
as supersymmetries of the deformed M2-brane.  It follows that in the
decoupling limit, the dual gauge theory on the 3-dimensional boundary
of AdS$_4$ has ${\cal N}=3$ supersymmetry.

    We also discussed the physical significance of this, and
previously-obtained resolved brane solutions, and analysed the
symmetry groups of the dual field theories.  Resolved brane solutions
lead to a breaking on conformal invariance in the dual field theories,
and this can provide a mechanism for confinement.  In fact the
mechanisms for the cases of dual field theories in $D=4$ and $D=3$ are
quite different.  In $D=4$, an additional fractional 5-brane flux is
needed for the brane-resolution, whilst in $D=3$ the breaking of the
conformal phase is caused by a perturbation of relevant operators,
associated with the pseudoscalar fields of the gauge theory in the
Higgs branch.

\appendix
\section{Ricci-flat manifolds in $D=4$, $6$, $7$ and $8$}

   In this appendix, we present a summary of known irreducible complete
non-compact Ricci-flat manifolds of cohomogeneity one in the dimensions
that are of relevance in string theory and M-theory.  In particular, we
shall discuss only those cases where the metric admits
covariantly-constant spinors, so that the associated resolved $p$-brane
solutions might be supersymmetric.  This means that the spaces must be
either Ricci-flat K\"ahler, or hyper-K\"ahler, or else with exceptional
holonomy $G_2$ in $D=7$ or Spin(7) in $D=8$.  In all the cases that we
shall consider, the metrics are of cohomogeneity one.  Most of the cases
are asymptotically conical, of the form $\R\times K$, where $K=G/H$ is a
homogeneous space whose metric approaches an Einstein metric of positive
curvature as $r$ tends to infinity.  The principal orbits $G/H$
typically degenerate at the ``centre,'' at the minimum value $r=r_0$ of
the radial coordinate.  The requirements of regularity of the manifold
at $r=r_0$ imply that the degeneration must be of the form $S^m\times
\wtd K$, where $S^m$ is a round sphere whose proper radius tends to zero
at the appropriate rate so that the manifold is locally of the form
$\R^{m+1}\times \wtd K$ near $r=r_0$.

\subsection{Ricci-flat K\"ahler metrics in $D=4$}

    First, we list the case of 4 dimensions. Here the only examples
within the class described above are K\"ahler (and hence
hyper-K\"ahler), with principal orbits that are locally $S^3$.  The most
relevant for our purposes, since it is asymptotically-conical, is the
Eguchi-Hanson solution \cite{egha}.  The global structure of this
manifold was first described in \cite{begipapo}, where it was shown that
the principal orbits are $RP^3\equiv S^3/\Z_2$ rather than $S^3$, and so
it provided the first example of an ALE space.

\cramp
\bigskip
\begin{center}
\begin{tabular}{|c|c|c|c|c|c|c|}\hline
\multicolumn{3}{|c|}{Principal Orbit}& Degenerate & Harmonic 
& Manifold & Name\\ \cline{1-3}
 $K$ & $G/H$& Isometry & Orbit $\wtd K$ & 2-Form? & & \\ \hline\hline
$S^3/\Z_2$ & $SU(2)$ & $U(2)$ & $S^2$ &  $L^2$ 
             & $T^*S^2$ &Eguchi-Hanson\\ \hline
$S^3$ & $SU(2)$ & $U(2)$ & -- & $L^2$  
       & $\R^4$ &Taub-NUT\\ \hline
$S^3/(\Z_2\times \Z_2)$ & $SU(2)$ & $SU(2)$ & $S^2$ & $L^2$ 
                     & $\wtd{T^*\RP^3}$  & Atiyah-Hitchin\\ \hline
\end{tabular}
\end{center}
\bigskip
\uncramp

\noindent{Table 3:} Complete non-compact Ricci-flat 4-manifolds with
covariantly-constant spinors \cite{egha,hawk,atihit}.  They are all
K\"ahler, with $SU(2)$ holonomy, and hence they are also
hyper-K\"ahler.  The $S^3$ principal orbits in Taub-NUT degenerate to
a point at the origin, just as in flat space.  (This, and its
higher-dimensional analogues, are discussed in appendix B.)  The tilde
in the expression $\wtd{T^*\RP^3}$ for the Atiyah-Hitchin manifold
denotes that it is the universal covering space of the co-tangent
bundle of $\RP^3$. Only the Eguchi-Hanson manifold is asymptotically
conical.

\subsection{Ricci-flat K\"ahler metrics in $D=6$}

    Next, we turn to $D=6$.  Here, there are two possibilities for the
principal orbits, namely $S^5$ or $T^{1,1}$.  These correspond to
$U(1)$ bundles over $\CP^2$ or $S^2\times S^2$ respectively.  
\bigskip

\begin{center}
\begin{tabular}{|c|c|c|c|c|c|}\hline
\multicolumn{3}{|c|}{Principal Orbit}& Degenerate & Harmonic 
& Manifold\\ \cline{1-3}
$K$ & $G/H$ & Isometry & Orbit $\wtd K$ & 3-Form? & \\ \hline\hline
$S^5/\Z_3$ 
    & $\fft{SU(3)}{SU(2)}$ & $U(3)$ & $\CP^2$ & Irreg. & $\C\ltimes \CP^2$ 
                               \\ \hline
$T^{1,1}$ & $\fft{SO(4)}{SO(2)}$ & $SO(4)\times U(1)$ & 
                    $S^2\times S^2$ & Irreg. 
       & $\C\ltimes S^2\times S^2$ \\ \hline
$T^{1,1}$ & $\fft{SO(4)}{SO(2)}$ & $SO(4)\times U(1)$ & $S^2$ & Irreg. 
                    & $\C^2\ltimes S^2$ \\ \hline
$T^{1,1}$ & $\fft{SO(4)}{SO(2)}$ & $SO(4)\times U(1)$ & $S^3$ & Reg, not $L^2$ 
                   & $T^*S^3$ \\ \hline
\end{tabular}
\end{center}
\bigskip

\noindent{Table 4:} Complete non-compact Ricci-flat 6-manifolds.
These are all K\"ahler, and so they have $SU(3)$ holonomy and admit two
covariantly-constant spinors..  The notation $X \ltimes Y$ is used to
signify a bundle over $Y$ with fibre $X$.  In the final row, $T^*S^3$
denotes the co-tangent bundle of $S^3$.  Metrics for the first two
examples are obtained in \cite{berber,pagpop}, and more general such
metrics for the second example are obtained in \cite{tz2}.  The metric
for the third example is in \cite{candel,tz1}, and for the fourth in
\cite{candel,sten,klebstra}.

\subsection{Ricci-flat metrics of $G_2$ holonomy in $D=7$}

    In $D=7$ the only irreducible manifolds within the class we are
listing here are ones with exceptional $G_2$ holonomy.  Three of these
are known explicitly.
\bigskip

\begin{center}
\begin{tabular}{|c|c|c|c|c|c|}\hline
\multicolumn{3}{|c|}{Principal Orbit}& Degenerate & Harmonic 
& Manifold\\ \cline{1-3}
$K$ & $G/H$ & Isometry &Orbit $\wtd K$ & 3-Form? & \\ \hline\hline
$\CP^3$ & $\fft{SO(5)}{SO(3)\times SO(2)}$ & $SO(5)\times SO(3)$ 
      & $S^4 $ & $L^2$ & $\R^3\ltimes S^4$ \\ \hline
Flag$_6$ & $\fft{SU(3)}{U(1)\times U(1)}$ & $SU(3)\times SU(2)$ 
      & $\CP^2$ & $L^2$ 
       & $\R^3\ltimes \CP^2$ \\ \hline
$S^3\times S^3$ & $\fft{SO(4)\times SO(3)}{SO(3)}$ & $SO(4)\times
SO(3)$  & $S^3$ & Reg, not $L^2$ 
                    & $\R^4\ltimes S^3$ \\ \hline
\end{tabular}
\end{center}
\bigskip

\noindent{Table 5:} Complete non-compact Ricci-flat 7-manifolds, which
all have exceptional $G_2$ holonomy.  Flag$_6$ denotes the 6-dimensional
flag manifold $SU(3)/(U(1)\times U(1))$.  The metrics are all
asymptotically conical, with principal orbits that approach Einstein
metrics on $\CP^3$, Flag$_6$ and $S^3\times S^3$.  The $\CP^3$ and
Flag$_6$ Einstein metrics are ``squashed,'' rather than the standard
K\"ahler ones. The Einstein metric on $S^3\times S^3$ is again
``squashed,'' and is not the standard direct-product metric.  The
metrics for all three examples are obtained in \cite{brysal,gibpagpop}.
Topologically, the manifolds can be described as the bundle of self-dual
2-forms over $S^4$ or $\CP^2$, and the spin bundle over $S^3$,
respectively.

\bigskip\bigskip

   For the sake of completeness, we shall present here the metrics,
and the superpotentials from which the Ricci-flat solutions can be derived,
for the three $G_2$ examples. We shall describe them in the order
listed in Table 5.  In fact the relevant equations are the same for
both the $\R^3$ bundle over $S^4$ and the $\R^3$ bundle over $\CP^2$.
The metrics can be written in the form \cite{gibpagpop}
%%%%%
\be
d s_7^2 =  dt^2  + c^2\,
(d\mu^i + \ep_{ijk}\, A^j\, \mu^k)^2 + a^2\, ds_4^2\,,
\ee
%%%%%
where $\mu^i\, \mu^i=1$ ($i=1,2,3$) and $A^i$ is the $SU(2)$
connection of the 4-dimensional quaternionic K\"ahler Einstein manifold with
metric $ds_4^2$.  This can be chosen to be $S^4$ or $\CP^2$.    Defining
a new radial coordinate $\eta$ by $dt = a^4\, c^2\, d\eta$, the
Ricci-flat conditions can be derived from the Lagrangian $L=T-V$,
together with the constraint $T+V=0$, where
%%%%%
\bea
T &\equiv & \ft12 g_{ij}\, {\a^i}'\, {\a^j}' = 
2 {\gamma'}^2 + 16 \a'\, \gamma' + 12\, {\a'}^2 \,,\nn\\
V &=&  -2 e^{2\gamma + 8 \a} - 12 e^{4\gamma +
6 \a} + 2 e^{6\gamma + 4 \a}\,,
\eea
%%%%%
where $a=e^\a$, $c=e^\gamma$ and a prime denotes $d/d\eta$.  We find
that the potential $V$ can be expressed in terms of a superpotential
$W$, such that $V=-\ft12 g^{ij}\, (\del W/\del\a^i)\, (\del
W/\del\a^j)$, with
%%%%%
\be
W = 4 e^{\gamma+4\a} + 4 e^{3\gamma + 2\a}\,.
\ee
%%%%%

   The first-order equations $d\a^i/d\eta = \ft12 g^{ij}\, \del
W/\del\a^j$, re-expressed in terms of the original radial variable
$t$, give
%%%%%
\be
\dot a= \fft{c}{a}\,,\qquad \dot c = 1- \fft{c^2}{a^2}\,.
\ee
%%%%%
The solution, after a further change of radial variable, gives the
complete Ricci-flat metric on the $\R^3$ bundle over $S^4$ or $\CP^2$
that was obtained in \cite{brysal,gibpagpop}:
%%%%%
\be
d\hat s_7^2 = \fft{dr^2}{1- r^{-4}} + \ft14 r^2\, (1 -r^{-4})\, (d\mu^i +
\ep_{ijk}\, A^j\, \mu^k)^2 + \ft12 r^2\, ds_4^2\,.
\ee
%%%%%

   For the $\R^4$ bundle over $S^3$, the form of the metric is
%%%%%
\be
ds_7^2 =  dt^2  + c^2\, (\nu_1^2 + \nu_2^2+\nu_3^2)
   + a^2\, (\sigma_1^2+\sigma_2^2+\sigma_3^2)\,,
\ee
%%%%% 
where $\nu_i=\Sigma_i-\ft12\sigma_i$, and the quantities $\sigma_i$
and $\Sigma_i$ are left-invariant 1-forms on two copies of $SU(2)$.  
In terms of a new radial variable $\eta$ defined by $dt=a^3\, c^3\,
d\eta$, the Ricci-flat conditions can be derived from the Lagrangian
$L=T-V$ with constraint $T+V=0$, where
%%%%%
\bea
T &=&  3 {\gamma'}^2 + 3 {\a'}^2  + 9 \a'\, \gamma' \,,\nn\\
V &=&  -\ft34 e^{4\gamma +
6\a} - \ft34 e^{6\gamma + 4\a} +\ft3{64} e^{8\gamma+ 2\a}\,.
\eea
%%%%%
We find that $V$ can be obtained from a superpotential, given by
%%%%%
\be
W= \ft32 e^{2\gamma + 3\a} + \ft38 e^{4\gamma + \a}\,,
\ee
%%%%% 

    The first-order equations following from this superpotential,
written in terms of $t$, are therefore
%%%%%
\be
\dot a = \fft{c}{4a}\,,\qquad \dot c = \ft12 - \fft{c^2}{8 a^2}\,.
\ee
%%%%%
The solution of these equations, in terms of a new radial variable
$r$, gives the complete Ricci-flat metric on the $\R^4$ bundle over
$S^3$ that was obtained in \cite{brysal,gibpagpop}:
%%%%%
\be
ds_7^2 = \fft{dr^2}{1 - r^{-3} }
+\ft19 r^2\, (1 - r^{-3})\, 
(\nu_1^2+\nu_2^2+\nu_3^2) + \ft1{12} r^2\,
(\sigma_1^2+\sigma_2^2 + \sigma_3^2)\,.
\ee
%%%%%  

\subsection{Ricci-flat metrics in $D=8$}

    In $D=8$ there are three kinds of irreducible manifolds within the
class we are listing here, with exceptional holonomy Spin(7),
Ricci-flat K\"ahler holonomy $SU(4)$, or hyper-K\"ahler holonomy
$Sp(2)\equiv $ Spin(5).  They admit 1, 2 or 3 covariantly-constant
spinors respectively. 
\bigskip

\begin{center}
\begin{tabular}{|c|c|c|c|c|c|c|}\hline
\multicolumn{3}{|c|}{Principal Orbit}& Degenerate & Harmonic 
& Manifold & Holonomy\\ \cline{1-3}
$K$ & $G/H$ & Isometry &Orbit $\wtd K$ & 4-Form? & & \\ \hline\hline
$S^7$ & $\fft{SO(5)}{SO(3)}$ & $\st{SO(5)\times SO(3)}$ 
      & $S^4 $ & $L^2$ & $\R^4\ltimes S^4$ 
            &Spin(7) \\ \hline\hline
$S^7/\Z_4$ & $\fft{SU(4)}{SU(3)}$ & $\st{U(4)}$ 
      & $\CP^3$ & $L^2$ 
       & $\C\ltimes \CP^3$ & $SU(4)$  \\ \hline
$N^{1,1}$ & $\fft{SU(3)}{U(1)}$ & $\st{U(3)}$ 
         & Flag$_6$ & $L^2$ 
            & $\C\ltimes \wtd K$ & $SU(4)$   \\ \hline
$M^{2,3}$ & $\fft{SU(3)\times SU(2)}{SU(2)\times U(1)} $ 
          & $\st{SU(3)\times U(2)}$ 
         & $S^2\times \CP^2$ & $L^2$ 
& $\C\ltimes \wtd K$ & $SU(4)$  \\ \hline 
$Q^{1,1,1}$ & $\fft{SU(2)\times SU(2)\times SU(2)}{U(1)\times U(1)}$ 
   & $\st{SU(2)^3\times U(1)}$ & $(S^2)^3$ 
 & $L^2$ & $\C\ltimes \wtd K$& 
    $SU(4)$    \\ \hline
$Q^{1,1,1}$ & $\fft{SU(2)\times SU(2)\times SU(2)}{U(1)\times U(1)}$ 
      & $\st{SU(2)^3\times U(1)}$ & $(S^2)^2$ 
 & $L^2$ & $\C^2\ltimes \wtd K$& 
    $SU(4)$  \\ \hline 
$V_{5,2}$ & $\fft{SO(5)}{SO(3)}$ & $\st{SO(5)\times U(1)}$ &
$S^4$ & $L^2$ & $T^* S^4$ & $SU(4)$ \\ \hline\hline
$N^{1,1}$ & $\fft{SU(3)}{U(1)}$ & $\st{U(3)}$ 
     & $\CP^2$ & $L^2$ &
     $T^*\CP^2$ & Spin(5) \\ \hline
\end{tabular}
\end{center}
\bigskip

\noindent{Table 6:} Complete non-compact Ricci-flat 8-manifolds. The
example with Spin(7) holonomy is obtained in \cite{brysal,gibpagpop}.
The first four K\"ahler examples are obtained in \cite{berber,pagpop},
and more general metrics for the third of these are obtained in
\cite{cglp1}, and for the fourth in \cite{cglp1,cglp2}.  The fifth
K\"ahler example is obtained in \cite{cglp1,cglp2}, and the final
K\"ahler example is contained in the construction of \cite{sten}, and
obtained explicitly in \cite{cglp1}.  Note that the embeddings of
$SO(3)$ in the descriptions of $S^7$ and $V_{5,2}$ as $SO(5)/SO(3)$ are
such that the {\bf 5} of $SO(5)$ decomposes to {\bf 1} + {\bf 2} + {\bf
2} and {\bf 1} + {\bf 1} + {\bf 3} respectively.  The hyper-K\"ahler
Calabi metric is obtained in \cite{calabi}, and a useful further
discussion is to be found in \cite{danswa}.  The fully explicit metric,
in a form that is most useful for our present purposes, is obtained in
the present paper.

\bigskip\bigskip

    For completeness, we shall present the metric, superpotential and 
solution for the complete Ricci-flat metric of Spin(7) holonomy on the
$\R^4$ bundle over $S^4$.  The metric is of the form
%%%%%
\be
ds_8^2 = dt^2 +  c^2 \, (\nu_1^2 + \nu_2^2 + \nu_3^2) 
+  a^2\, d\Omega_4^2\,,\label{qkmetric}
\ee
%%%%% 
where $\nu_i=\sigma_i-A^i$, and $A^i$ denotes the potential
for the $SU(2)$ Yang-Mills instanton over the unit 4-sphere with
metric $d\Omega_4^2$.  In terms of a new radial variable $\eta$,
defined by $dt=a^4\, c^3\, d\eta$, the Ricci-flat conditions can be
derived from the Lagrangian $L=T-V$, with the constraint $T+V=0$,
where
%%%%%
\bea
T &=& 6{\gamma'}^2 + 24 \a'\, \gamma' + 12\, {\a'}^2 \,,\nn\\
V &=&  -\ft32 e^{4\gamma + 8 \a} - 12 e^{6\gamma +
6\a} + 3  e^{8\gamma + 4 \a}\,, 
\eea
%%%%%
One finds that $V$ can be derived from a superpotential\footnote{This
superpotential was obtained in \cite{kanyas}.  A system of first-order
equations that gives rise to the metric of \cite{brysal,gibpagpop} 
was also obtained in \cite{baflke}.}
%%%%%
\be
W= 3 e^{2\gamma + 4\a} + 6 e^{4\gamma+2\a}\,.
\ee
%%%%%

    In terms of the original radial variable $t$, the first-order
equations following from this superpotential are
%%%%%
\be
\dot a = \fft{3 c}{2a}\,,\qquad \dot c = \ft12 -\fft{c^2}{a^2}\,.
\label{spin7fo}
\ee
%%%%%
The solution of these equations yields the complete Ricci-flat metric
of Spin(7) holonomy obtained in \cite{brysal,gibpagpop}:
%%%%%
\be
ds_8^2 = \fft{dr^2}{1- r^{-10/3}} + \ft9{100} r^2\, 
(1 -r^{-10/3})\, (\sigma_i-A^i)^2 + \ft9{20} r^2\, 
d\Omega_4^2\,,\label{spin7mets4}
\ee
%%%%%
Note that the first-order equations (\ref{spin7fo}) are the same, after 
trivial rescalings, as those obtained for the 8-metrics of Spin(7) holonomy
in section 4.1.

\section{Higher-dimensional generalisations of Taub-NUT} 

    Although the Taub-NUT metric \cite{hawk} and its
higher-dimensional generalisations \cite{baisbate} are not
asymptotically conical, it is useful to gather the results for these
metrics here, since they provide further explicit examples of complete
Ricci-flat metrics. The construction gives generalisations in all even
dimensions, but only the 4-dimensional Taub-NUT metric itself is K\"ahler.
(See also \cite{awacha} for a recent discussion of some
higher-dimensional metrics of the Taub-NUT type.)

    Consider the general class of complex lines-bundle metrics
%%%%%
\be
d\hat s^2 = h^2\, dr^2 + c^2\, \sigma^2 + a^2\, d\Sigma_m^2
\ee
%%%%%
where $d\Sigma_m^2$ is the Fubini-Study metric on the unit $\CP^m$
(with cosmological constant $\lambda=2m+2$, where $R_{ij}=\lambda\,
\delta_{ij}$).  The 1-form $\sigma$ is given by
%%%%%
\be
\sigma= dz + A\,,
\ee
%%%%%
where $dA= 2 J$, and $J$ is the K\"ahler form on $\CP^m$.

    In the vielbein basis $\hat e^0=h\, dr$, $\hat e^{\td 0} = c\,
\sigma$, $\hat e^a = a\, e^a$, the Ricci tensor of $d\hat s^2$ is
given by
%%%%%
\bea
\hat R_{00} &=& - \fft{n}{h\, a}\, \Big(\fft{a'}{h}\Big)' - \fft1{h\,
c}\, \Big(\fft{c'}{h}\Big)'\,,\nn\\
\hat R_{\td0\td0} &=& -\fft{n\, a'\, c'}{h^2\, a\, c} - \fft{1}{h\,
c}\, \Big(\fft{c'}{h}\Big)' + \fft{n\, c^2}{
a^4}\,,\label{ricci}\\
\hat R_{ij} &=& -\Big[ \fft1{a\, h}\, \Big(\fft{a'}{h}\Big)' +
\fft{a'\, c'}{h^2\, a\, c} + \fft{(n-1)\, {a'}^2}{h^2\, a^2}
-\fft{\lambda}{a^2} + \fft{2 c^2}{a^4}\Big]\, \delta_{ij}\,,\nn
\eea
%%%%%
where $n=2m$ is the real dimension of the $\CP^m$ base space. 

   Making the coordinate gauge choice $h\, c=1$, we find that the
combination $\hat R_{00}-\hat R_{\td0\td0}=0$ of the Ricci-flat
conditions implies that $a'' + a^{-3}=0$, which gives
%%%%%
\be
a^2 = r^2-1\,.
\ee
%%%%%
The equation $\hat R_{00}=0$ then implies that $c^2=k\, f$, where $k$
is a constant and $f$ satisfies the equation
%%%%%
\be
(r^2-1)^2\, f'' + n\, r\, (r^2-1)\, f' - 2n\, f=0\,.
\ee
%%%%%
We require the solution that is non-singular at $r=1$.  This is given
by
%%%%%
\be
f= r\, (r^2-1)^{-m}\, \int_1^r s^{-2}\, (s^2-1)^{m}\, ds\,,\label{fsol}
\ee
%%%%%
(recalling that $n=2m$), which, in terms of hypergeometric functions,
is
%%%%%
\be
f= (-1)^m\, r\, (r^2-1)^{-m}\, 
\Big( \fft{2^{2m}\, (m!)^2}{(2m)!} -\fft1{r}\,\,
 _2F_1[-\ft12, -m, \ft12, r^{2}]\Big)\,.
\label{fsol2}
\ee
%%%%%
(For any integer $m$, $f$ reduces to a rational function of $r$.  The
first few cases are presented below.)
It can then be verified that the remaining Einstein equation $\hat
R_{ij}=0$ is satisfied provided that the constant $k$ in $c^2=k\, f$
is chosen to be $k=2m+2$.  It is useful to note from (\ref{fsol}) that
$f$ satisfies the first-order equation
%%%%%
\be
r\, (r^2-1)\, f' + [1 + (n-1)\, r^2]\, f = r^2-1\,.
\ee
%%%%%

   Thus we arrive at the Ricci-flat metric
%%%%%
\be
d\hat s^2 = \fft{dr^2}{(2m+2)\, f} + (2m+2)\, f\, \sigma^2 + (r^2-1)\,
d\Sigma_m^2\,.\label{gentn}
\ee
%%%%%
For $r$ close to 1, if we write $r=1+\ft12 \rho^2$, it follows from
(\ref{fsol}) that
%%%%%
\be
f \sim \fft{\rho^2}{2m+2}\,,
\ee
%%%%%
and the metric approaches
%%%%%
\be
d\hat s^2 = d\rho^2 + \rho^2\, ( \sigma^2 + d\Sigma_m^2)\,.
\ee
%%%%%
Since $dA=2J$, it follows that $\sigma^2 + d\Sigma_m^2$ is precisely
the metric on the unit $(2m+1)$-sphere, and so near $r=1$ the metric
(\ref{gentn}) smoothly approaches the origin of $\R^{2m+2}$ written in
spherical polar coordinates.   At large $r$ we have
%%%%%
\be
f \sim \fft1{2m-1}\,,
\ee
%%%%%
and so the metric is asymptotically flat with the cylindrical geometry
$\R^{2m+1}\times S^1$. 

   When $m=1$, the resulting 4-dimensional metric is precisely
Taub-NUT.  For larger $m$, we get higher-dimensional generalisations.
The first few examples are
%%%%%
\crampest
\bea
d\hat s_4^2 & =& \fft{r+1}{4(r-1)}\, dr^2 + \fft{4(r-1)}{r+1}\,
\sigma^2 + (r^2-1)\, d\Sigma_1^2\,,\nn\\
d\hat s_6^2 & =& \fft{(r+1)^2}{2(r-1)(r+3)}\, dr^2 +
 \fft{2(r-1)(r+3)}{(r+1)^2}\,
\sigma^2 + (r^2-1)\, d\Sigma_2^2\,,\\
d\hat s_8^2 & =& \fft{5(r+1)^3}{8(r-1)(r^2+4r+5)}\, dr^2
  + \fft{8(r-1)(r^2+4r+5)}{5 (r+1)^3}\,
\sigma^2 + (r^2-1)\, d\Sigma_3^2\,,\nn\\
d\hat s_{10}^2 & =& \fft{7(r+1)^4}{2(r-1)(5r^3+25r^2+47r+35)}\, dr^2
 + \fft{2(r-1)(5r^3+25r^2+47r+35)}{7(r+1)^4}\,
\sigma^2 + (r^2-1)\, d\Sigma_4^2\,.\nn
\eea
\uncramp
%%%%% 

   The generalised Taub-NUT metrics are not K\"ahler, except for the
special case $n=2m=2$ of the original 4-dimensional example.  One way to
see that they are not in general K\"ahler is to note that if they
were, then, being Ricci flat, they would admit two 
covariantly-constant spinors $\eta$. 
These would have to satisfy the integrability conditions 
$R_{abcd}\, \Gamma_{cd}\, \eta=0$. To show that these cannot be
satisfied for $m\ge2$ it suffices to look at the 
components $\Theta_{0i}$ of the curvature 2-form:
%%%%%
\be
\Theta_{0i} =k\, Q_1\, e^0\wedge e^i + k\, Q_2\, J_{ij}\, e^{\td 0}\wedge
e^j\,,
\ee
%%%%%
where
%%%%%
\be
Q_1 = \fft{1-r^2 + [3+ (n-1)\, r^2]\,f }{2(r^2-1)^2}\,,\qquad
Q_2 = \fft{1-r^2 + [1+(n+1)\, r^2]\, f}{2r\, (r^2-1)^2}\,.
\ee
%%%%%
It is evident that the associated integrability condition is
%%%%%
\be
Q_1\, \Gamma_{0i}\, \eta + Q_2\, J_{ij}\, \Gamma_{\td 0 j}\, \eta=0\,,
\ee
%%%%%
and that this can only be satisfied by a non-vanishing $\eta$ if
$Q_1=\pm Q_2$.  It is straightforward to check from the expression
(\ref{fsol2}) that this can only occur if $n=2$.

\section{A more general $T^*\CP^2$ construction in $D=8$}

   In our construction of the Calabi metrics in section 2, we took the 
$U(1)$ generator within the coset to be $\lambda=L_1{}^1-L_2{}^2$,
while the $U(1)$ generator corresponding to the combination
$Q=L_1{}^1+L_2{}^2$ was taken to lie outside the coset.  It is
possible to take a more general embedding of the $U(1)$ generators, in
which we have $\lambda$ within the coset and $Q$ outside the coset
given by
%%%%%
\bea
\lambda &=& (L_1{}^1-L_2{}^2)\, \cos\delta + (L_1{}^1+L_2{}^2)\,
\sin\delta\,,\nn\\
Q &=& -(L_1{}^1-L_2{}^2)\, \sin\delta + (L_1{}^1+L_2{}^2)\,
\cos\delta\,,
\eea
%%%%%
where $\delta$ is a fixed angle that specifies the embedding.  It can
take a discrete infinity of possible values, and the case discussed in
section 2 that leads to the Calabi hyper-K\"ahler metrics corresponds
to $\delta=0$.  

    We shall just present results for the more general $\delta\ne0$
embeddings in $D=8$ here, since again we find that there exists a
superpotential leading to first-order equations that imply Spin(7)
holonomy.  Specifically, we find that the superpotential
(\ref{spin7spot}) now becomes
%%%%%
\bea
W&=& 4 a^3\, b\, c\, f + 4 a\, b^3\, c\, f + 4 a\, b\, c^3\, f - 
4a^2\, b^2\, f^2\, \cos\delta \nn\\
&&-2\sqrt2\, a^2\, c^2\, f^2\, \sin(\delta-\ft14\pi) 
+2\sqrt2\, b^2\, c^2\, f^2\, \cos(\delta-\ft14\pi) \,.
\eea
%%%%%
Correspondingly, the first-order equations (\ref{fo}) generalise to 
%%%%%
\bea
\dot\a &=& \fft{b^2+c^2-a^2}{a\, b\, c} - 
      \fft{\sqrt2\, f\, \cos\td\delta}{a^2}\,,\nn\\
\dot\beta &=& \fft{a^2+c^2-b^2}{a\, b\, c} + 
              \fft{\sqrt2\, f\, \sin\td\delta}{b^2}\,,\nn\\
\dot\gamma &=& \fft{a^2+b^2-c^2}{a\, b\, c} +
      \fft{\sqrt2\, f\,(\cos\td\delta -\sin\td\delta) }{c^2}\,,\nn\\
\dot\sigma &=& - \fft{\sqrt2\, f\, (\cos\td\delta-\sin\td\delta)}{c^2}
   + \fft{\sqrt2\, f\, \cos\td\delta }{a^2} -
\fft{\sqrt2\, f\, \sin\td\delta}{b^2}\,,\label{fo2}
\eea
%%%%%
where we have defined $\td\delta= \delta-\ft14\pi$.  

\section{Complex and quaternionic conifolds}

\subsection{Complex conifolds}

   Let us first briefly review the construction of complex conifolds,
which describe the cones over $SO(n+2)/SO(n)$.  These all admit smooth
resolutions to give the Stenzel metric of dimension $D=2n+2$ on
$T^*S^{n+1}$, the co-tangent bundle of $S^{n+1}$.  The case $n=1$
corresponds to Eguchi-Hanson and $n=2$ corresponds to the deformed
conifold of \cite{candel,klebstra} that was used in the construction
of the fractional D3-brane.  The case $n=3$ is an 8-dimensional
metric, which was used in \cite{cglp1} in order to construct a smooth
deformed M2-brane.  The conifold for $n=3$ was discussed in
\cite{klewit}.

   For arbitrary $n$, we generalise the discussion in
\cite{candel,klewit}, and consider complex coordinates $z_A$ on
$\C^{n+2}$, subject to the quadratic constraint
%%%%%
\be
\sum_A z_A^2=0\,.\label{ccon1}
\ee
%%%%%
The rays under which $z_A$ is identified with $\lambda\, z_A$ for all
$\lambda$ form a cone.  Fixing the radius to unity, without loss of
generality, we have also
%%%%%
\be
\sum_A |z_A|^2=1\,.\label{ccon2}
\ee
%%%%%
The quadric $\sum z_A^2$ in (\ref{ccon1}) is invariant under
$z_i\longrightarrow z_A'=A_{AB}\, z_B$ for complex matrices $A_{AB}$
satisfying $A^T\, A=\oneone$, \ie in $SO(n+2,\C)$, while (\ref{ccon2})
is invariant for $A^\dagger\, A=\oneone$, \ie under $U(n+2)$.  Writing
$A= P+ \im\, Q$ where $P$ and $Q$ are real, we see that from
$(A^T-A^\dagger)A=0$ we shall have $Q^T\, Q=0$, and hence by taking the
trace we get $Q_{AB}\, Q_{AB}=0$, which implies $Q=0$.  The intersection
of $SO(n+2,\C)$ and $U(n+2)$ is therefore $SO(n+2)$, and this acts
transitively on the base of the cone defined by (\ref{ccon1}) and
(\ref{ccon2}).  A fiducial point on this manifold, for example
%%%%%
\be
z_1= \fft1{\sqrt2}\,,\qquad z_2=\fft{\im}{\sqrt2}\,,\qquad
z_3=z_4=\cdots z_{n+2}=0\,,
\ee
%%%%%
is stabilised by $SO(n)$, acting on the last $n$ of the $z_A$.  Thus
the base of the cone is the coset space $SO(n+2)/SO(n)$.  There is
also an additional $U(1)$ factor in the isometry group, giving
$SO(n+2(\times U(1)$ in total.  This arises because the fact that
$\sum_A z_A^2$ vanishes in (\ref{ccon1}) allows the additional
symmetry transformation $z_A\longrightarrow e^{\im\, \gamma}\, z_A$,
which is also an invariance of (\ref{ccon2}).

   The Einstein metric on the coset $SO(n+2)/SO(n)$ that forms the
base of the cone can be constructed in terms of the $z_A$ coordinates,
as  
%%%%%
\be
ds^2 = \fft{n}{n+1}\, \Big[
\sum_A |dz_A|^2 - k^2\, \Big|\sum_A \bar z_A\, dz_A\Big|^2 \Big]
\,,\label{stencon}
\ee
%%%%%
where $k$ is a constant ``squashing parameter'' whose value we shall
determine presently.  To show this, let $Z$ be the column vector of $z_A$
coordinates, so $Z^T=(z_1,z_2,\ldots , z_{n+2})$, and let $Z_0$ be the
fiducial point defined above, \ie $Z_0^T = (1,\im,0,\ldots,0)/\sqrt2$.
Then as we saw above, we may express the general point $Z$ as
%%%%%
\be
Z= P\, Z_0\,,
\ee
%%%%%
where $P$ is a (real) orthogonal matrix in $SO(n+2)$.  We shall
therefore have
%%%%%
\bea
\sum_A |dz_A|^2 &=& dZ^\dagger\, dZ = Z_0^\dagger\, dP^T\, dP\, Z_0 
                  = Z_0^\dagger\, K^T\, K\, Z_0\,,\nn\\
\sum_A \bar z_A \, dz_A &=& Z_0^\dagger\, K\, Z_0\,,
\eea
%%%%%
where we have defined the 1-form $K$ in the Lie algebra of $SO(n+2)$ 
by $K\equiv P^T\, dP$.  This may be expanded in terms of the $SO(n+2)$ 
generators $T_{AB}$ as 
%%%%%
\be
K = \ft12 L_{AB}\, T_{AB}\,,
\ee
%%%%%
where $L_{AB}$ are left-invariant 1-forms on the group $SO(n+2)$.  

    If we define the basis $(n+2)$-vectors $X_A$, by
$X_A^T=(0,0,\ldots,0,1,0,\ldots,0)$, where there is a zero in every
component except for a 1 at the $A$'th position, then for $SO(n+2)$
generators $T_{AB}$ in the fundamental representation we shall have
%%%%%
\be
T_{AB}\, X_C =\delta_{BC}\, X_A - \delta_{AC}\, X_B\,.
\ee
%%%%%
Writing the fiducial point as $Z_0=(X_1+ \im\, X_2)/\sqrt2$, it
is now an elementary exercise to show that
%%%%%
\be
Z_0^\dagger\, K^T\, K\, Z_0 = \ft12 \sigma_i^2 +  \ft12 \td\sigma_i^2 
                             + \nu^2\,,\qquad
 Z_0^\dagger\, K\, Z_0 = \im\, \nu
\ee
%%%%%
where we have defined
%%%%%
\be
\nu\equiv L_{12}\,,\qquad \sigma_i\equiv L_{1i}\,,\qquad 
    \td\sigma_i \equiv L_{2i}\,,\qquad 3\le i \le n+2\,.
\ee
%%%%%
Note that $\nu$, $\sigma_i$ and $\td\sigma_i$ are the left-invariant
1-forms on the coset $SO(n+2)/SO(n)$ that we introduced in
\cite{cglp1} in our construction of the Stenzel metrics on
$T^*S^{n+1}$.  Thus we see that the proposed metric (\ref{stencon}) on
the $SO(n+2)/SO(n)$ coset that forms the base of the cone can be
written as
%%%%%
\be
ds^2 = \fft{n}{n+1}\, \Big[
\ft12 \sigma_i^2 + \ft12 \td\sigma_i^2 + (1-k^2)\, \nu^2\Big]\,.
\label{stencon2}
\ee
%%%%%

    This metric may be compared with the form of the Stenzel metric 
in $D=2n+2$ dimensions \cite{sten}, in the form obtained in \cite{cglp1};
%%%%%
\be
d\hat s^2 = c^2\, dr^2 + c^2\, \nu^2 + a^2\, \sigma_i^2 + b^2\,
\td\sigma_i^2\,,\label{stenmet}
\ee
%%%%% 
where 
%%%%%
\be
a^2= R^{1/(n+1)}\, \coth r\,,\quad
b^2 = R^{1/(n+1)}\, \tanh r\,,\quad
c^2= \fft1{n+1}\, R^{-n/(n+1)}\, (\sinh
2r)^n\,,
\ee
%%%%% 
and 
%%%%%
\be
R(r) \equiv  \int_0^r (\sinh 2u)^n\, du\,.\label{rdef}
\ee
%%%%%
From this, it is easy to see that at large $r$ the Stenzel metric
approaches 
%%%%%
\be
d\hat s^2 = d\rho^2 + \rho^2\, ds_0^2\,,\label{stenbig}
\ee
%%%%%
where
%%%%%%
\be
ds_0^2 = \fft{n}{n+1}\,
\Big[\ft12 \sigma_i^2 + \ft12 \td\sigma_i^2 + \fft{n}{n+1}\, \nu^2\Big]\,.
\label{stenorb}
\ee
%%%%%
This metric, which is necessarily Einstein,\footnote{Since $d\hat s^2$
in (\ref{stenmet}), and its asymptotic form (\ref{stenbig}), is Ricci
flat, it follows that $ds_0^2$ in (\ref{stenbig}) must be Einstein, with
$R_{ab} = 2n\, g_{ab}$.} is the required form for the metric on
$SO(n+2)/SO(n)$ on the base of the cone.  Comparing with
(\ref{stencon2}), we see that it agrees provided that we choose
%%%%%
\be
k^2 = \fft{1}{n+1}\,.\label{kval}
\ee
%%%%%
The metric (\ref{stencon}) for the case $n=2$ was obtained in
\cite{candel}.  This case is rather special, in that the
transitively-acting group is $SO(4)$ which factorises as $SU(2)\times
SU(2)$, and so the principal orbits can be described as the 5-manifold
$T^{1,1}$ which is a $U(1)$ bundle over $S^2\times S^2$, whose metric
was given in \cite{pagpopwh}.

     To summarise, we have shown that the Stenzel metrics, which are
regular and asymptotic to cones over $SO(n+2)/SO(n)$, have the structure
at large radius given by (\ref{stenbig}), with the metric on the
principal orbits approaching (\ref{stenorb}).  This Einstein metric on
$SO(n+2)/SO(n)$ is the one that is needed in the conifold construction
of the cone over $SO(n+2)/SO(n)$, and we have shown that it can be
written in terms of the conifold coordinates $z_A$ as (\ref{stencon}),
with the constant $k$ given by (\ref{kval}).  It should be emphasised,
therefore, that the Einstein metric on $SO(n+2)/SO(n)$ 
is {\it not} obtained simply by restricting the flat metric $d\bar z_A
\, dz_A$ on $\C^{n+2}$ by the quadratic constraints (\ref{ccon1}) and
(\ref{ccon2}).  Rather, we must subtract the appropriate multiple 
of $|\bar z_A\, dz_A|^2$, as given by (\ref{stencon}) and
(\ref{kval}).  This amounts to a homogeneous squashing of the fibres
in $SO(n+2)/SO(n)$, viewed as a $U(1)$ bundle over the real
Grassmannian $SO(n+2)/(SO(n)\times SO(2))$.

\subsection{Quaternionic conifolds and the hyper-K\"ahler quotient}

\subsubsection{Quaternionic conifolds}

    We saw in the previous subsection how the complex conifold
construction of \cite{candel} for $D=6$, and \cite{klewit} for $D=8$,
generalises to arbitrary complex dimensions, and that the Stenzel
metrics on $T^*S^{n+1}$ can be viewed as the resolutions of the
singular metrics on the cones.  Here, we carry out a similar analysis
for the analogous quaternionic conifolds.  The construction, known as
the hyper-K\"ahler quotient, shows explicitly how the Calabi metrics
on $T^*CP^{n+1}$ can be viewed as the resolutions of the singular
metrics on the quaternionic cones.  Some care is needed in the
construction, on account of the non-commutativity of the quaternions.
It is useful to write a quaternion $q$ in terms of two complex
quantities $u$ and $v$:
%%%%%
\be
q = \pmatrix{u & -\bar v\cr
             v & \bar u}\,.\label{comqua}
\ee
%%%%%

    We begin by considering $(n+2)$ quaternions $q_A$ on $\H^{\, n+2}$,
subject to the quadratic constraint
%%%%%
\be
\sum_A q_A^T\, q_A =0\,,\label{qcon1}
\ee
%%%%%
where $q_A^T$ denotes transposition of the $2\times 2$ matrix
introduced in (\ref{comqua}).  The rays where $q_A$ is identified with
$q_A\, \lambda$, with $\lambda$ any quaternion, define a
cone,\footnote{Note that unlike the complex conifold discussed
previously, here one has to distinguish between left-multiplication
and right-multiplication by quaternions.  In fact (\ref{qcon1}) is
slightly too weak a condition to specify the cone, since it is also
invariant under the left-multiplication $q_A\longrightarrow \kappa\,
q_A$, where $\kappa$ is a quaternion satisfying $\kappa^T\, \kappa=1$.
In fact by writing (\ref{qcon1}) in terms of the complex components 
$u_a$ and $v_a$ as in (\ref{comqua}), we see that it implies 
$u_A^2 + v_A^2=0$ and $\bar u_A\, v_A - \bar v_A\, u_A=0$, which is just 
three real conditions, and not four as one might have expected.
As we shall see later, the factoring out of the residual $U(1)$
implies that we should perform a Hopf reduction on the fibres
corresponding to this $U(1)$ coordinate.} 
and fixing the radius to unity gives the
intersection with the unit $(4n+7)$-sphere
%%%%%
\be
\sum_i q_A^\dagger\, q_A =1\,.\label{qcon2}
\ee
%%%%%
Under transformations $q_A\longrightarrow q_A'= A_{AB}\, q_B$, with
%%%%%
\be
q_A \equiv  \pmatrix{u_A & -\bar v_A\cr
             v_A & \bar u_A}\,,\qquad 
A_{AB} = \pmatrix{\a_{AB} & -\bar\beta_{AB}\cr
                  \beta_{AB} & \bar\a_{AB}}\,,\label{qtouv}
\ee
%%%%%
we see that $q_A^T\, q_A$ is invariant for matrices $A$ in
$SO(n+2,\H)= SO^*(2n+4)$, while $q_A^\dagger\, q_A$ is invariant for
matrices $A$ in $Sp(n+2)=USp(2n+4)$.  In terms of the complex
components $\a_{AB}$ and $\beta_{AB}$, these conditions are
%%%%%
\bea
SO(n+2,\H):&& \a^T\, \a + \beta^T\, \beta = \oneone\,,\qquad
  \a^\dagger\, \beta - \beta^\dagger\, \a=0\,,\nn\\
Sp(n+2):&& \a^\dagger\, \a + \beta^\dagger\, \beta = \oneone\,,\qquad
  \a^T\, \beta - \beta^T\, \a=0\,.\label{mcon}
\eea
%%%%%  

   The intersection of $SO(n+2,\H)$ and $Sp(n+2)$ can be seen by
defining $\a=\a_R + \im\, \a_I$, $\beta= \beta_R+\im\, \beta_I$, where
the matrices $\a_R$, $\a_I$, $\beta_R$ and $\beta_I$ are all real.
Taking differences of the two lines in (\ref{mcon}), and then taking
the imaginary parts, we learn that $\a_I^T\, \a_I=0$ and $\beta_I^T\,
\beta_I=0$, and hence $\a_I=0$, $\beta_I=0$.  Having learned that the
matrices $\a$ and $\beta$ are real, it is easy to see that by defining
the complex matrix $S\equiv \a+\im\, \beta$, we shall have
%%%%%
\bea
S^\dagger\, S &=& (\a^T - \im\, \beta^T)(\a+\im\, \beta)\,,\nn\\
      &=& \a^T\, \a + \beta^T\, \beta + \im\, (\a^T\, \beta
-\beta^T\, \a)\,,\label{ssun}
\eea
%%%%% 
and so the remaining conditions in (\ref{mcon}) are satisfied if and
only if $S$ is unitary, $S^\dagger\, S=\oneone$.  The $U(1)$ factor in
$U(n+2)=SU(n+2)\times U(1)$ is irrelevant here, and so this shows that
the manifold defined by (\ref{qcon1}) and (\ref{qcon2}) is invariant
under $SU(n+2)$, which acts transitively on the base of the cone.  As in
the complex case there is in fact an additional $U(1)$ isometry, coming
from the fact that the right-hand side in (\ref{qcon1}) is zero,
implying that the $q_A$ can all undergo the phase-scaling transformation
$q_A\longrightarrow e^{\im\, \gamma}\, q_A$, which also leaves
(\ref{qcon2}) invariant.  Thus the total isometry group is $U(n+2)$.

   The coset structure can be seen by choosing a fiducial point on the
rays satisfying (\ref{qcon1}) and (\ref{qcon2}), such as
%%%%%
\be
q_1 = \fft1{\sqrt2}\, \pmatrix{1 & 0\cr 0 & 1}\,,\qquad
q_2 = \fft1{\sqrt2}\, \pmatrix{0 & \im \cr \im & 0}\,,\qquad
q_3 =q_4 =\cdots = q_{n+2}=0\,.\label{quatfid}
\ee
%%%%%
This has $U(n)$ as the stability subgroup contained within the
transitively-acting $SU(n+2)$ obtained above, 
consisting of matrices $S$ built as above, but with
%%%%%
\be
\a_{11}=\a_{22}=1\,,\qquad
\a_{12}=\a_{21}=\beta_{12}=\beta_{21}=\beta_{11} =\beta_{22}=0\,.
\ee
%%%%%
Thus the base of the cone is the homogeneous manifold $SU(n+2)/U(n)$. 

   The coset $SU(n+2)/U(n)$ is precisely the one that we used in
section 2, for the principal orbits of the manifolds of cohomogeneity
one that led to the hyper-K\"ahler Calabi metrics on $T^*\CP^{n+1}$.
Thus these smooth manifolds are resolutions of the quaternionic cones
in $\H^{\, n+2}$ defined by (\ref{qcon1}).

   As in the previous discussion for the complex conifolds, here too
we can give an explicit expression for the Einstein metric on the
coset space $SU(n+2)/U(n)$ that forms the base of the 
cone, in terms of the quaternionic conifold coordinates.  To do this,
it is advantageous to introduce further notation as follows.  We
define complex column vectors $U$ and $V$ by $U^T=(u_1,u_2,\ldots,
u_{n+2})$ and $V^T=(v_1,v_2,\ldots,v_{n+2})$ respectively, and then
define the quaternionic column vector $Z$ by
%%%%%
\be
Z \equiv U + \jm\, V\,,
\ee
%%%%%
where $\jm$ is an imaginary quaternion unit, which anticommutes with
$\im$ used in the construction of the complex quantities, with $\im\,
\jm = -\jm\, \im = \km$, {\it etc}.  It is easy to see that the
original conditions (\ref{qcon1}) and (\ref{qcon2}) are nothing but
%%%%%
\be
Z^T\, Z=0\,,\qquad Z^\dagger\, Z = 1\,.
\ee
%%%%%

   We also now define the special unitary matrix $P$ by
%%%%%
\be
P\equiv \a + \beta\, \jm\,,
\ee
%%%%%
where as above $\a$ and $\beta$ are real matrices satisfying
(\ref{mcon}).\footnote{{\it Qua} representation of $U(n+2)$, $P$ is
equivalent to the special unitary matrix $S$ introduced above equation
(\ref{ssun}).  Since we are now changing to a description of the
quaternions in terms of the three imaginary units $(\im,\jm,\km)$
rather than in terms of complex $2\times 2$ matrices, we need to use
the $\jm$ unit in the construction of $P$.}   
It is easy to see that the action of $U(n+2)$ on the
original quaternionic coordinates $q_A$, which are related to $u_A$
and $v_A$ as in (\ref{qtouv}), is now expressible as
%%%%%
\be
Z\longrightarrow Z'= P\, Z= \wtd P\, e^{\jm\, \psi}\, Z\,,
\ee
%%%%%
where we have extracted the phase from $P$ in $e^{\jm\, \psi}$, so
that $\det \wtd P=1$. 
The fiducial point (\ref{quatfid}) is now given by
%%%%%
\be
Z_0 = U_0 +  \jm\, V_0 = \ft1{\sqrt2}\, X_1 - \ft1{\sqrt2}\, \km\,
X_2\,,\label{fidpoint}
\ee
%%%%%
where $X_A$ is the same set of basis vectors as in the complex case
(zero except for a 1 at $A$'th position).  

   Following analogous steps to those for the complex conifold, we now
introduce the 1-form $K\equiv P^\dagger\, dP$ in the Lie algebra of
$U(n+2)$;
%%%%%
\be
K \equiv P^\dagger\, dP = \wtd P^\dagger\, d\wtd P + \jm\, d\psi =
\jm\, L_A{}^B\, T_B{}^A + \jm\, d\psi\,,\label{kfromp}
\ee
%%%%%
where $T_B{}^A$ are the (hermitean) generators of $SU(n+2)$
(constructed using $\jm$ as the complex unit),
satisfying $(T_B{}^A)^\dagger = T_A{}^B$.  In the fundamental
representation we shall have
%%%%%
\be
T_A{}^B\, X_C = \delta^B_C\, X_A - \fft1{n+2}\, 
\delta_A{}^B\, X_C\,.\label{tonx}
\ee
%%%%
(The second term, which is a trace subtraction, will not enter in the
subsequent calculations since $L_A{}^A=0$.) Noting that we  have
$L_A{}^B \, \km = \km\, L_B{}^A$, it is now a straightforward exercise
to derive the following:
%%%%%
\be
dZ^\dagger\, dZ = Z_0^\dagger\, K^\dagger\, K\, Z_0 = \ft12(
   \sigma_{1\a}^2 + \sigma_{2\a}^2 + \Sigma_{1\a}^2 + \Sigma_{2\a}^2) 
   + \nu_1^2+\nu_2^2 + \ft14\lambda^2 + \ft14(d\psi + \ft12 Q)^2\,,
\label{dzres}\nn
\ee
%%%%%
where the real 1-forms $(\sigma_{1\a}\,
\sigma_{2\a},\Sigma_{1\a},\Sigma_{2\a},\nu_1,\nu_2\,\lambda,Q)$
appearing here are as defined in section 2 of the paper.  Note in
particular that $Q=L_1{}^1 + L_2{}^2$ is the $U(1)$ generator that
lies {\it outside} the coset.

    In our proof of the relation between the conditions
(\ref{mcon}) and $S^\dagger\, S=1$ (or $P^\dagger\, P=1$) in
(\ref{ssun}), we noted that the phase of $S$ (or $P$) is irrelevant,
since it factors out in $S^\dagger\, S=1$.  Therefore in our
construction of the metric on conifold, we should project orthogonally
to the orbit of the $U(1)$ fibres $\psi$ that parameterise this phase
(see (\ref{kfromp})).  Thus we see from
(\ref{dzres}) that the metric $dZ^\dagger\, dZ$ induced from the
original flat metric on $\H^{\, n+2}$, after this projection
perpendicular to the $U(1)$ Hopf $U(1)$ fibres, is 
%%%%%
\be
ds^2 = dZ^\dagger\, dZ\Big|_{\rm perp} = 
 \ft12(
   \sigma_{1\a}^2 + \sigma_{2\a}^2 + \Sigma_{1\a}^2 + \Sigma_{2\a}^2) 
   + \nu_1^2+\nu_2^2 + \ft14 \lambda^2\,.\label{afterp}
\ee
%%%%%

    We may now compare this with the asymptotic form of the metric on
the $SU(n+2)/U(n)$ principal orbits in the hyper-K\"ahler Calabi
metrics, which are given by (\ref{calabimetric}).  At large distance
we have
%%%%%
\bea
d\hat s^2 &=& dr^2 + r^2\, ds_0^2\,,\nn\\
 &=& dr^2 + r^2\, \Big[ \ft12(
   \sigma_{1\a}^2 + \sigma_{2\a}^2 + \Sigma_{1\a}^2 + \Sigma_{2\a}^2) 
   + \nu_1^2+\nu_2^2 + \ft14 \lambda^2 \Big]\,.
\eea
%%%%%
Thus the metric on the $SU(n+2)/U(n)$ base of the
asymptotically-conical large-distance limit of the Calabi metrics is
precisely the same as the one we obtained in (\ref{afterp}) from our
quaternionic conifold construction above.  An interesting feature is
that in this hyper-K\"ahler case we are finding that the induced
metric coming simply from $dZ^\dagger\, dZ$, followed by a Hopf
reduction on the $U(1)$ fibres associated with the transformation
$q_A\longrightarrow \kappa\, q_Q$, yields the required Einstein metric
on the $SU(n+2)/U(n)$ base of the cone.  By contrast, in the complex
conifold case that we discussed previously, we instead obtained the required
Einstein metric on the $SO(n+2)/SO(n)$ base of the cone by subtracting 
an appropriate multiple of $|Z^\dagger\, dZ|^2$ from the metric directly 
induced from $|dZ^\dagger\, dZ|^2$.  The difference between
the two cases is a reflection of the more rigid structure of
hyper-K\"ahler metrics, in comparison to Ricci-flat K\"ahler metrics.

\subsubsection{Hyper-K\"ahler quotient}

   In this section we make contact with the hyper-K\"ahler quotient
construction of the Calabi metrics, as described, for example, in
\cite{gibryc}.  We present a very simple rederivation of the Calabi
metrics in the notation of this paper, using the hyper-K\"ahler 
quotient construction.

   The construction proceeds as follows.  Let us define a column
vector of quaternions $W$ by
%%%%%
\be
W\equiv r\, Z\,,
\ee
%%%%%
where, as in section (D.2.1), $Z$ is normalised so that $Z^\dagger\,
Z=1$, and so we have $W^\dagger\, W=r^2$.  We now deform the original
cone metric, and modify (\ref{qcon1}), which would imply $W^T\, W=0$,
so that instead
%%%%%
\be
W^T\, W = - \ell^2\,,
\ee
%%%%%
where $\ell$ is a real constant.\footnote{ The quaternion $q$ in
\cite{gibryc} is related to our $W$, and the complex components $u$
and $v$ in (\ref{comqua}) are related to the $q$ in \cite{gibryc} by
$q=r\, (u - v\, \jm)$ (the r\^oles of left and right multiplication in
\cite{gibryc} are interchanged in comparison to the conventions we have
adopted here).  The $U(1)$ action in our discussion is equivalent to
transforming the $q$ of \cite{gibryc} according to
$q\longrightarrow q\, e^{\im\, t}$.}  We can achieve this, which
implies $Z^T\, Z = -\ell^2/r^2$, while still
maintaining $Z^\dagger\, Z=1$,  by changing our
fiducial point (\ref{fidpoint}) so that now $Z_0$ is given by
%%%%%
\be
Z_0 = \ft1{\sqrt2}\, \Big[ \Big(1-\fft{\ell^2}{r^2}\Big)^{1/2}\, X_1 
       - \km\,\Big(1+\fft{\ell^2}{r^2}\Big)^{1/2}\, X_2  \Big]\,.
\ee
%%%%%
It is now a simple exercise to calculate the metric
%%%%%
\be
d\hat s^2 \equiv dW^\dagger\, dW = dr^2 + r^2\, dZ^\dagger\, dZ
= dr^2 + r^2\, Z_0^\dagger\, K^\dagger\, K\, Z_0 + r^2\,
dZ_0^\dagger\, dZ_0\,,
\ee
%%%%%
giving
%%%%%
\bea
d\hat s^2 &=& \Big(1-\fft{\ell^4}{r^4}\Big)^{-1}\, dr^2  
+ \ft14 r^2\, \Big(1-\fft{\ell^4}{r^4}\Big) \, \lambda^2 + r^2\,
(\nu_1^2 + \nu_2^2) \nn\\
&&+ \ft12 (r^2-\ell^2)\, 
(\sigma_{1\a}^2 + \sigma_{2\a}^2) + \ft12 (r^2+\ell^2)\, 
(\Sigma_{1\a}^2 + \Sigma_{2\a}^2) \nn\\
&& + r^2\, \Big( d\psi + \ft12 Q - \fft{\ell^2}{2 r^2}\, \lambda
\Big)^2\,.\label{prequot}
\eea
%%%%%
We see that after projecting orthogonally to the vector
$\del/\del\psi$ along the $U(1)$ fibres, we end up with hyper-K\"ahler 
Calabi metric in precisely the form we derived in section 2, with
scale parameter $\ell=1$.  Note that the final term in (\ref{prequot})
is nothing but $r^2\, (d\psi -2A_1)^2$, where $A_1$ is the potential
for the K\"ahler form $J_1$ obtained in (\ref{3ares}).

\section*{Acknowledgements}

We should like to thank Bobby Acharya, Andrew Dancer, Nigel Hitchin
and Matt Strassler for discussions.  M.C. is supported in part by DOE
grant DE-FG02-95ER40893 and NATO grant 976951; H.L.~is supported in
full by DOE grant DE-FG02-95ER40899; C.N.P.~is supported in part by
DOE DE-FG03-95ER40917.  The work of M.C., G.W.G. and C.N.P. was
supported in part by the programme {\it Supergravity, Superstrings and
M-theory} of the Centre \'Emile Borel of the Institut Henri
Poincar\'e, Paris (UMS 839-CNRS/UPMC).

\end{document}